\UseRawInputEncoding

\documentclass[aps,pre,amsmath,amssymb,floatfix,reprint,citeautoscript,noeprint,superscriptaddress,twocolumn]{revtex4-2}
\usepackage{lineno}

\usepackage{dsfont}
\usepackage{lipsum} 
\usepackage{bibentry}
\usepackage[english]{babel}
\usepackage[dvipsnames]{xcolor}
\usepackage{graphicx}
\usepackage[caption=false]{subfig} 
\usepackage{amsmath,amssymb,bm}
\usepackage[version=3]{mhchem}
\usepackage{verbatim}
\usepackage{multirow}
\usepackage{dcolumn}
\usepackage{float}
\usepackage{nicefrac}
\usepackage{siunitx}
\usepackage{booktabs}
\usepackage{chemformula}
\usepackage{wrapfig}
\usepackage{enumitem}  
\usepackage{transparent}
\usepackage[colorlinks,allcolors=black,citecolor=blue,urlcolor=blue]{hyperref}
\usepackage{tcolorbox}
\usepackage{relsize}
\hypersetup{
    citecolor=Turquoise,
    colorlinks=true,
    linkcolor=black,    
    urlcolor=Turquoise,
    }
\usepackage{titlesec}

\DeclareSIUnit[number-unit-product = {\,}]{\amu}{amu}
\DeclareSIUnit[number-unit-product = {\,}]{\kJmol}{\kilo\joule\per\mol}
\DeclareSIUnit[number-unit-product = {\,}]{\Nsm}{\newton\second\per\meter\cubed}
\DeclareSIUnit[number-unit-product = {\,}]{\THz}{\tera\hertz}
\DeclareSIUnit[number-unit-product = {\,}]{\meV}{\milli\electronvolt}
\DeclareSIUnit[number-unit-product = {\,}]{\cal}{cal}

\newcommand{\tbf}[1]{\textbf{#1}}

\newcommand{\mrm}[1]{\mathrm{#1}}

\usepackage{sansmathfonts}
\usepackage[justification=centerlast, font={sf}, labelfont={bf}]{caption}
\begin{document}

\title{Molecular structure breaks universality in screening behavior of electrolyte solutions}

\author{Anna T. Bui}
\email{annatbui@stanford.edu}
\affiliation{Yusuf Hamied Department of Chemistry, University of
  Cambridge, Lensfield Road, Cambridge, CB2 1EW, United Kingdom}
\affiliation{Department of Chemistry, Durham University, South Road,
  Durham, DH1 3LE, United Kingdom}
\affiliation{Department of Chemistry, Stanford University, California 94305, USA}

\author{Connie J. Fairchild}
\affiliation{Yusuf Hamied Department of Chemistry, University of
  Cambridge, Lensfield Road, Cambridge, CB2 1EW, United Kingdom}

\author{Dominic P. W. Thomas}
\affiliation{Department of Chemistry, Durham University, South Road,
  Durham, DH1 3LE, United Kingdom}
  
\author{Stephen J. Cox}
\email{stephen.j.cox@durham.ac.uk}
\affiliation{Department of Chemistry, Durham University, South Road,
  Durham, DH1 3LE, United Kingdom}

\date{\today}


\begin{abstract}
  Electrostatic screening is one of the most fundamental
    phenomena in electrolyte solutions. While dilute electrolytes are
    described by the universal Debye--H\"uckel theory, recent
    experiments have reported an apparently universal scaling law for
    screening lengths in concentrated electrolytes, fueling ongoing
    debate over anomalous underscreening. Yet, liquid state theory has
    been unable to determine whether molecular details of the solvent,
    the ions, or both should preserve or break such
    universality. Using machine-learned classical density functional
    theory, we compute numerically exact decay lengths across a
    hierarchy of electrolyte models of increasing molecular
    complexity. We find that molecular details qualitatively reshape
    screening: solvent attraction generates a new monotonic decay
    mode, solvent polarity produces re-entrant behavior, and ionic
    structure alone can drastically alter screening-length
    exponents. Consequently, no single scaling law describes screening
    universally across these molecularly distinct electrolytes. Our
    results challenge equilibrium interpretations of experimentally
    observed universality in concentrated electrolytes.
\end{abstract}

\maketitle

\section{INTRODUCTION}

\begin{figure*}[t]
  \includegraphics[width=\linewidth]{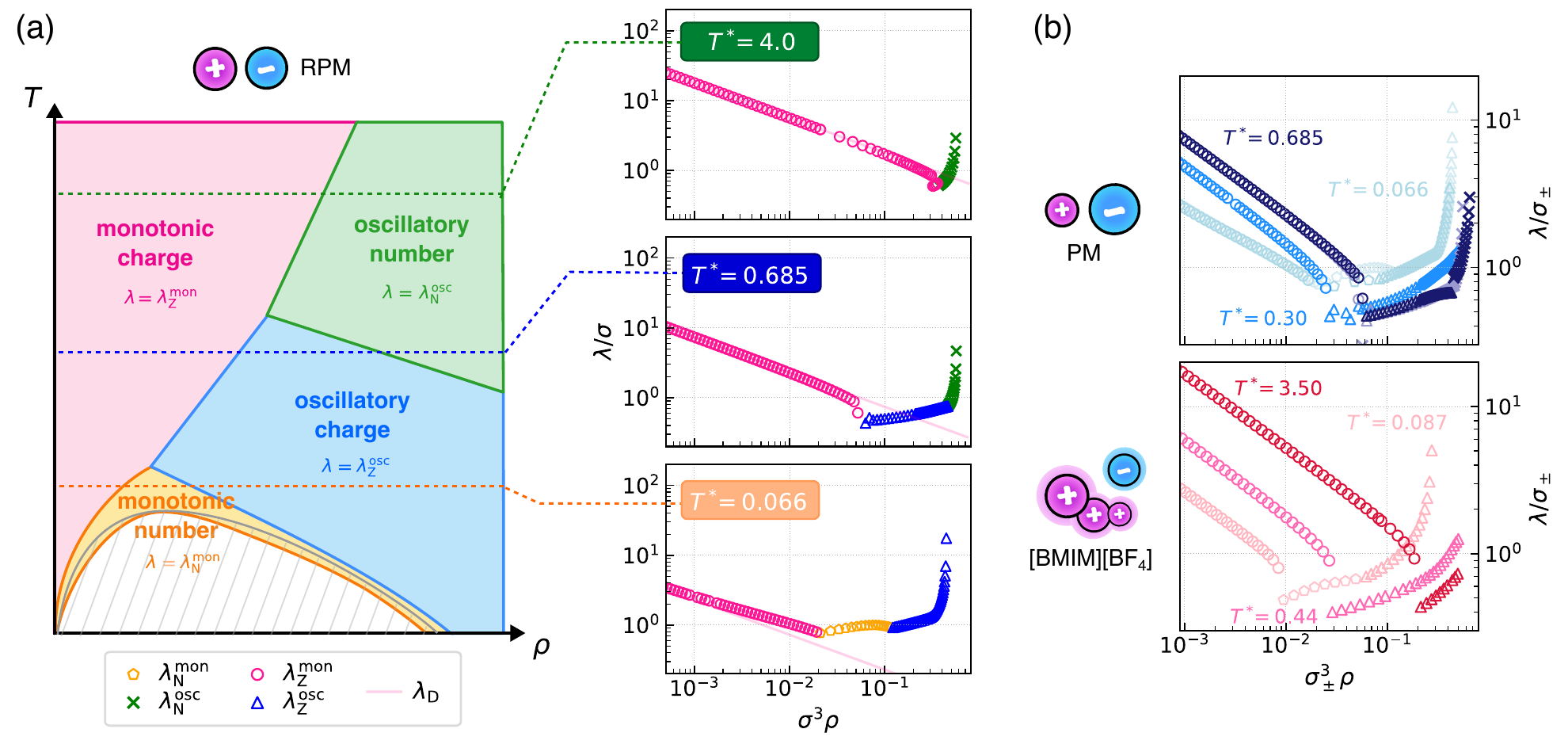}
  \caption{\tbf{Decay length regimes of solvent-free ionic fluids.}
    (a) Schematic diagram in the temperature--density $(T,\rho)$ plane
    identifying the four asymptotic decay regimes of the RPM,
    reconstructed from Ref.~\cite{leote_de_carvalho_decay_1999}. Decay
    lengths computed from the neural cDFT are overlaid along the three
    isotherms marked by dashed lines.  The solid gray line marks the
    binodal.  The dominant decay length transitions between monotonic
    charge, oscillatory charge, oscillatory number and monotonic
    number as density and temperature vary. (b) Decay length for an
    asymmetric primitive model (PM, top) and for a coarse-grained
    model of the ionic liquid [BMIM][BF$_4$] (bottom), confirming that
    the broad picture established in (a) persists under ionic size
    asymmetry and molecular detail.}
\label{fig1}
\end{figure*}

Over large separations, the effective interaction between charged
species in solution is governed by the electrostatic screening
length. Understanding how the properties of a solution determine
the screening length is, therefore, central to controlling a range of
bulk and interfacial phenomena, such as colloidal stability
\cite{Leunissen2005}, energy storage in electric double layer
supercapacitors \cite{Salanne2016}, and electrochemical actuation in
soft materials \cite{Kim2016}. At low ionic strength, electrostatic
screening exhibits a largely universal character: Debye--H\"uckel
theory \cite{Debye1923} eschews molecular detail in favor of generic
solution properties such as ionic strength ($I$), solvent dielectric
constant ($\epsilon$), and temperature ($T$). Specifically, the
electrostatic potential generated by a charged species in an
electrolyte is exponentially screened over the Debye length,
\begin{equation}
  \lambda_{\mathrm{D}}=\left(\frac{\epsilon\epsilon_0 k_{\mathrm{B}}T}{2e^2I}\right)^{1/2},
\end{equation} 
where $\epsilon_0$ is the permittivity of free space, $e$ is the
  elementary charge, and $k_{\rm B}$ is the Boltzmann
constant. The ionic strength is determined by the number
  densities ($\{\rho_\nu\}$) and valencies ($\{z_\nu\}$) of the ions,
  $2I = \sum_\nu z_\nu^2\rho_\nu$. Its simple, universal structure has
  made Debye--H\"uckel theory a cornerstone of our understanding of
  dilute electrolytes. \cite{Levin2002,  Buell2013, Pluharova2017, Gonella2021,Rehl2022, Wu2022}

Beyond the dilute regime, no comparable universal theoretical
description has emerged.
Increasing ion density introduces steric packing and correlations that
drive a crossover from the monotonic decay predicted by Debye-H\"uckel
theory to a damped oscillatory decay; this ``Kirkwood transition''
marks the onset of alternating charge layers in the fluid
\cite{Kirkwood1936}.  Surface force measurements have long provided
experimental access to interactions beyond the dilute regime,
revealing oscillatory hydration forces in aqueous electrolytes
\cite{Pashley1984} and long-range, monotonically decaying forces in
room-temperature ionic liquids \cite{Gebbie2013, Gebbie2015,
  Hjalmarsson2017}. More recently, measurements across a wide range of
concentrated electrolytes and room-temperature ionic liquids
\cite{Smith2016, Lee2017} have reported that the screening length of
the oscillatory mode, $\lambda^{\rm osc}$, increases steeply with
concentration, following an apparently universal scaling law of
$(\lambda^{\rm
  osc}/\lambda_{\mathrm{D}})\sim(\sigma/\lambda_{\mathrm{D}})^\eta$,
where $\sigma$ is the ion size.  This steep scaling law with
$\eta\approx3$ has attracted considerable, still unresolved, debate;
some studies report no evidence of anomalous underscreening
\cite{Kumar2022}, some observe lower exponents \cite{Gaddam2019,
  Rondepierre2025, Dinpajooh2024, Robertson2023, Elliott2024}, while
others have pointed to non-equilibrium effects \cite{Cross2026,
  tilger2026colloidalprobeatomicforce}.

More advanced liquid state theories that incorporate short-ranged
ion--ion correlations successfully predict a crossover from monotonic
to oscillatory decay with increasing ionic strength, yet have failed
to reproduce the reported experimental scaling \cite{Ennis1995,
  Ulander2001, Rotenberg2018, Cats2021}. These theories rely on
approximate descriptions of both fluid correlations and the underlying
molecular interactions; in most studies, electrolytes are represented
using ``primitive'' models, in which ions are treated as charged hard
spheres embedded in a structureless dielectric continuum. Such
primitive models have been studied extensively, with the general
conclusion that the experimentally reported underscreening reflects
physics beyond that contained in this simple representation of
electrolyte interactions \cite{Cats2021}.

Molecular simulations offer a natural alternative route, but
extracting the true asymptotic decay length is challenging. Both
spatial and temporal finite-size effects make it difficult to resolve
the long, weak tail of correlation functions from statistical noise;
such effects become particularly severe at high concentrations, where
the reported underscreening corresponds to increasingly long-ranged
correlations. Nevertheless, simulations have found no evidence of the
anomalously large screening lengths reported experimentally
\cite{Coles2020, Zeman2020}. With neither molecular simulation nor
current liquid-state theories able to conclusively reproduce the
reported scaling of electrostatic screening at high ionic strength, it
remains an open question whether a universal screening law exists in
concentrated electrolytes at all.

In recent years, machine learning has offered a way to bypass many of
the challenges faced by traditional liquid-state theories.
Specifically, in classical density functional theory (cDFT), neural
density functionals can be learned directly from simulation data,
yielding highly accurate approximations to exact free-energy
functionals without invoking uncontrolled approximations
\cite{Sammuller2023}. This approach has proven successful in capturing
a diverse range of phenomena, including liquid--vapor coexistence,
liquid--liquid phase separation, dielectrocapillarity, and azeotropy
\cite{Sammuller2025, bui2025dielectro, zhou2026, Robitschko2025,
  Bui2026}. Particularly relevant to the present work, neural cDFT has
also enabled asymptotic analyses of fluid correlations. For example,
neural cDFT has been used to determine the Fisher--Widom line, which
marks the crossover between monotonic and oscillatory decay of density
correlations, in both the Lennard-Jones fluid \cite{Sammuller2025} and
molecular carbon dioxide \cite{Bui2026}. We have also extended neural
cDFT to ionic fluids \cite{Bui2025prl, bui2025hyperlmft}, where it
accurately reproduces the structure and thermodynamics of both bulk
and inhomogeneous systems. However, our previous studies were limited
to primitive electrolyte models, and no asymptotic analysis of
correlation functions was performed.

In this article, we take advantage of neural cDFT for ionic systems to
directly assess how molecular-scale complexity influences
electrostatic screening across a diverse range of systems. We adopt a
systematic approach that enables efficient mapping of screening
behavior throughout the temperature--density--concentration
manifold. We first demonstrate that neural cDFT faithfully reproduces
known behavior in primitive electrolytes and investigate asymptotic
correlations in an ionic liquid (Fig.~\ref{fig1}), before
systematically introducing explicit solvent degrees of freedom.  This
allows us to isolate the influence of solvent size, dispersion
interactions, and polarity on electrostatic screening. Finally, we
examine single-component plasma models to identify the minimal
physical ingredients underlying nontrivial screening behavior in
highly concentrated electrolytes.

Our central finding is that molecular details matter. We find that the
properties of both solvent and ions qualitatively reshape screening
throughout the concentrated regime: solvent attraction drives the
emergence of a new monotonic decay mode, dipolar solvents produce
re-entrant behavior, and ionic architecture can strongly alter the
apparent scaling of the screening length. Screening lengths comparable
in magnitude to experimental measurements also emerge, but at higher
concentrations, indicating that their large magnitude alone is not
anomalous. Collectively, these results challenge the notion of a
universal screening law in concentrated electrolytes. Instead,
electrostatic screening emerges as a consequence of molecular
architecture, opening opportunities to engineer screening behavior
through the molecular design of solvents and ions.

\section{RESULTS AND DISCUSSION}

\subsection{Decay length of solvent-free ionic fluids}

Even without a solvent, the asymptotic decay of correlations in ionic
fluids is already remarkably rich. Depending on the thermodynamic
conditions, the total correlation function between species $\nu$ and
$\gamma$ decays either monotonically,
$ rh_{\nu\gamma}(r) \sim \exp(-\alpha_0 r),$ or through damped
oscillations,
$rh_{\nu\gamma}(r) \sim \exp(-\alpha_0
r)\cos(\alpha_1r-\theta_{\nu\gamma})$. In both cases, the decay length
is $1/\alpha_0$, with $\alpha_0>0$, and, when oscillatory, the
wavelength is $2\pi/|\alpha_1|$.  The dominant decay can be associated
with either charge or number density correlations
\cite{LeotedeCarvalho1994, Attard1993}.  These asymptotic forms follow
from the leading poles of $\hat{h}_{\nu\gamma}$, obtained from the
Ornstein--Zernike equation \cite{LeotedeCarvalho1994, Evans1994},
\begin{equation}
  \hat{h}_{\nu\gamma}(k) = \hat{c}_{\nu\gamma}(k) + \sum_{\delta}\rho_{\delta}\,\hat{c}_{\nu\delta}(k)\,\hat{h}_{\delta\gamma}(k),
\end{equation}
where $\hat{h}_{\nu\gamma}$ and $\hat{c}_{\nu\gamma}$ are the Fourier
transforms of the total correlation function and the two-body direct
correlation functions between species $\nu$ and $\gamma$,
respectively, and $\rho_\delta$ is the number density of species
$\delta$. For an equimolar binary electrolyte, the poles
$\alpha=\alpha_1+i\alpha_0$ satisfy
\begin{equation}
  \left[1 - \frac{\rho \hat{c}_{++}(\alpha)}{2}\right]
  \left[1 - \frac{\rho \hat{c}_{--}(\alpha)}{2}\right]
  - \frac{\rho^2 \hat{c}_{+-}^2(\alpha)}{4} = 0,
\end{equation}
where $\rho = \sum_\delta\rho_\delta$. Among the poles contributing to
a given correlation function, those with the smallest 
imaginary part $\alpha_0$ govern the asymptotic decay. A purely
imaginary leading pole produces monotonic exponential decay, whereas a
leading pair $\alpha=\pm\alpha_1+i\alpha_0$ produces damped
oscillations.

Due to the long-range nature of the Coulomb interaction, it is
necessary to resolve electrostatic contributions to $c_{\nu\gamma}$.
Unlike previous works that treat the full Coulomb potential explicitly
\cite{leote_de_carvalho_decay_1999}, the accuracy of neural cDFT for
ionic systems rests upon ideas from local molecular field theory
(LMFT), in which the Coulomb potential is split exactly into short-
and long-ranged contributions \cite{Rodgers2008}. We refer the reader
to Refs.~\cite{Archer2013, Gao2020, Bui2025prl, bui2025hyperlmft} for
detailed discussions on the connections between cDFT and LMFT.  Most
relevant here is that the LMFT decomposition leads to the splitting of
the direct correlation function
\begin{equation}
  c_{\nu\gamma}(r) = c_{\mathrm{R},\nu\gamma}(r)
- \frac{\beta e^2 z_\nu z_\gamma}{4\pi\epsilon\epsilon_0} \frac{\mathrm{erf}(\kappa r)}{r},
\label{eqn:LMFT-splitting}
\end{equation}
where $\beta=(k_{\mrm{B}}T)^{-1}$ and $\kappa^{-1}$ sets the
LMFT splitting length. Direct correlation functions of the
reference system, $c_{{\rm R},\nu\gamma}$, therefore account only for
short-ranged electrostatic correlations, while the long-ranged
component is treated at the mean-field level. Importantly, LMFT
provides a systematic route to improving upon the mean-field
approximation: increasing $\kappa^{-1}$ transfers progressively more
electrostatic correlations into the reference system. In the limit
$\kappa^{-1}\to0$, all electrostatic interactions are treated in a
mean-field fashion, recovering the standard approach
\cite{LeotedeCarvalho1994, leote_de_carvalho_decay_1999, Cats2021}.
The key advantage for the present work is that this short-ranged
reference system is amenable to neural cDFT.

To determine the character of the decay mode, it is convenient
to define
\begin{equation}
  \label{eqn:RPM-basis}
  \begin{aligned}
    c_{\mrm{NN}}(r) & = \tfrac{1}{4} \left[c_{++}(r) + c_{--}(r) + 2 c_{+-}(r) \right], \\
    c_{\mrm{ZZ}}(r) &=  \tfrac{1}{4} \left[ c_{++}(r) +  c_{--}(r)  - 2 c_{+-}(r) \right], \\
    c_{\mrm{NZ}}(r) &=  \tfrac{1}{4} \left[  c_{++}(r) -   c_{--}(r) \right]. 
  \end{aligned}
\end{equation}
For ions of equal and opposite charge, only $c_{\mrm{ZZ}}$ retains the
long-range Coulombic tail; number--number and number--charge
correlations are faithfully described by the reference system, i.e.,
$c_{\mrm{NN}} = c_{\mrm{R,NN}}$ and $c_{\mrm{NZ}} =
c_{\mrm{R,NZ}}$. Using the LMFT decomposition
(Eq.~\ref{eqn:LMFT-splitting}), the pole condition reads
\begin{equation}
  \begin{aligned}
    \left[1 - \rho \hat{c}_{\mrm{R, NN}}(\alpha)\right]
    \left[1 - \rho \hat{c}_{\mathrm{R},\mathrm{ZZ}}(\alpha)
    + \frac{\lambda_{\mathrm{D}}^{-2}}{\alpha^2}
    \exp\!\left(-\frac{\alpha^2}{4\kappa^2}\right)\right] \\
    - \rho^2 \hat{c}_{\mrm{R,NZ}}^2(\alpha)= 0,
    \label{eqn:pole-LMFT}
  \end{aligned}
\end{equation}
where $\lambda_{\mathrm{D}}$ is the Debye length. For a symmetric
  electrolyte, $c_{\mrm{R, NZ}}=0$, and the number and charge
density poles decouple, making explicit that the asymptotic decay can
be governed by either number or charge density correlations.

We obtain $c_{\mrm{R},\nu\gamma}$ directly from neural cDFT, avoiding
the need for additional analytical approximations. Following previous
studies \cite{Sammuller2023, Sammuller2025, Bui2025prl}, we train a
neural network to accurately represent
$c_{\mrm{R},\nu}^{(1)}(z;[\{\rho_{\nu}\}],T)$, the first functional
derivative of the excess Helmholtz free-energy functional of the
reference fluid, using grand-canonical Monte Carlo simulation
data. The two-body direct correlation functions
$c_{\mrm{R},\nu\gamma}(r;[\{\rho_{\nu}\}],T)$ are then obtained via
reverse-mode automatic differentiation and subsequent radial
projection (see Methods). For each phase point $(\rho, T)$, the roots
of Eq.~\ref{eqn:pole-LMFT} thus yield up to four characteristic decay
lengths, $1/\alpha_0$. The asymptotic decay length of the fluid is
given by
\begin{equation}
   \lambda = \max\left(\lambda^{\mrm{mon}}_{\mrm{N}}, \lambda^{\mrm{osc}}_{\mrm{N}}, \lambda^{\mrm{mon}}_{\mrm{Z}},  \lambda^{\mrm{osc}}_{\mrm{Z}}\right),
\end{equation}
from which the character of the decay---monotonic number, oscillatory
number, monotonic charge, or oscillatory charge---can be determined.

To validate the approach, we first consider the restricted primitive
model (RPM), in which cations and anions are modeled as hard spheres
of equal diameter $\sigma_+ = \sigma_- = \sigma$ carrying charges
$\pm z e$. Screening in the RPM has been widely studied, both with
liquid-state theory \cite{LeotedeCarvalho1994, Cats2021, Ennis1995}
and molecular simulations \cite{Hartel2023}.  In
Fig.~\ref{fig1}a we present decay lengths obtained along three
representative supercritical isotherms spanning a broad range of
reduced temperatures
$T^* = \sigma_{\pm}/\lambda_{\rm B}$, each corresponding to a
different valency-scaled Bjerrum length,
$\lambda_{\mathrm{B}}=\beta z^2e^2/(4\pi\epsilon\epsilon_0)$. Here,
$\sigma_{\pm}=(\sigma_++\sigma_-)/2$ is the mean ion diameter,
introduced to accommodate unequal ion sizes later.

At sufficiently low densities, the long-ranged nature of the Coulomb
interaction dominates asymptotic correlations, and Debye--H\"uckel
screening is observed at all temperatures investigated. As density
increases, the character of the leading pole is, however, sensitive to
temperature. At $T^\ast = 4.0$, thermal fluctuations suppress the
relative importance of electrostatic correlations, and the asymptotic
decay becomes governed by packing-induced number--density correlations
for $\rho\sigma^3\gtrsim 0.4$. Reducing temperature to
$T^\ast = 0.685$, we instead observe a decay-mode dominated by charge
density correlations at intermediate densities,
$0.05 \lesssim \rho\sigma^3\lesssim 0.5$, before the
oscillatory-number decay mode takes over.  Our results closely
reproduce the decay lengths and sequence of decay regimes and
transitions reported in Ref.~\onlinecite{Cats2021}, as shown in
Fig.~S1.

For the lowest temperature investigated, $T^\ast = 0.066$, the system
is not so far from the critical point. Here, the relative
strength of electrostatic interactions is such that even the monotonic
charge decay mode deviates from Debye--H\"uckel theory, with
$\lambda_{\rm Z}^{\rm mon} > \lambda_{\rm D}$ at densities well below
the crossover to a monotonic-number mode at
$\rho\sigma^3\approx 0.02$. Increasing density further, we observe a
transition to an oscillatory-charge mode at $\rho\sigma^3\approx
0.1$. Aside from acting to validate our approach against previous
work, these results highlight that even for the RPM, the nature of
asymptotic correlations is remarkably rich and depends qualitatively
on the thermodynamic conditions.

Relaxing the constraint of equal diameters, we next consider an
asymmetric primitive model (PM) with $\sigma_{-} = 4\sigma/3$ and
$\sigma_{+} = 2\sigma/3$ such that $\sigma_{\pm}=\sigma$. In contrast
to the RPM, $c_{\rm NZ}$ is non-zero, and charge and number poles no
longer fully decouple. Nevertheless, the PM's decay length behavior
remains nearly identical to that of the RPM, consistent with treating
size asymmetry as a perturbation in this regime \cite{Cats2021,
  LeotedeCarvalho1994}. As shown in Fig.~\ref{fig1}b, the charge poles
are essentially unchanged, while the number poles are only weakly
perturbed, with slightly reduced decay lengths relative to the
RPM. Overall, the topology of the decay length diagram is essentially
unchanged.

Moving beyond primitive models entirely, we also examined a
coarse-grained representation of the ionic liquid [BMIM][BF$_4$]
\cite{Merlet2012}. To obtain $c_{\nu\gamma}$ for such a molecularly
structured fluid, we employed the hyperdensity functional theory
formulation of neural cDFT, training a neural network to represent
$c_{\mrm{R},\nu}^{(1)}(z;[\{\rho_{\nu}\},\beta\phi],T)$, where
$\phi(z)$ is the external electrostatic potential
\cite{bui2025hyperlmft,bui2025dielectro}. For this ionic liquid,
$\sigma_{\pm}=4.445\,\mathrm{\AA}$ was taken as an effective diameter
based on the cation--anion contact distance. As our focus is on
  asymptotic decay, we obtain $c_{\rm ZZ}$ using
  Eq.~\ref{eqn:RPM-basis}, treating the molecular ions as point
  charges in the far field despite the cation's charge being
  distributed across multiple sites. Remarkably, despite quantitative
differences, the qualitative decay length behavior remains unchanged
from that observed in the RPM and asymmetric PM. Owing to low GCMC
acceptance rates, we did not probe the higher-density
regime. Nevertheless, given the close correspondence with the RPM and
asymmetric PM at lower densities, we anticipate that a transition to
an oscillatory-number decay mode would also occur at sufficiently high
density.

Overall, these solvent-free results reveal a robust four-regime
picture that is largely unaffected by ionic size asymmetry and
  persists even in a molecularly structured ionic liquid. This
provides a baseline against which we can now isolate the role of an
explicit solvent.

\subsection{Solvent effects add interplay between total density and concentration}

\begin{figure*}[t]
  \includegraphics[width=0.88\linewidth]{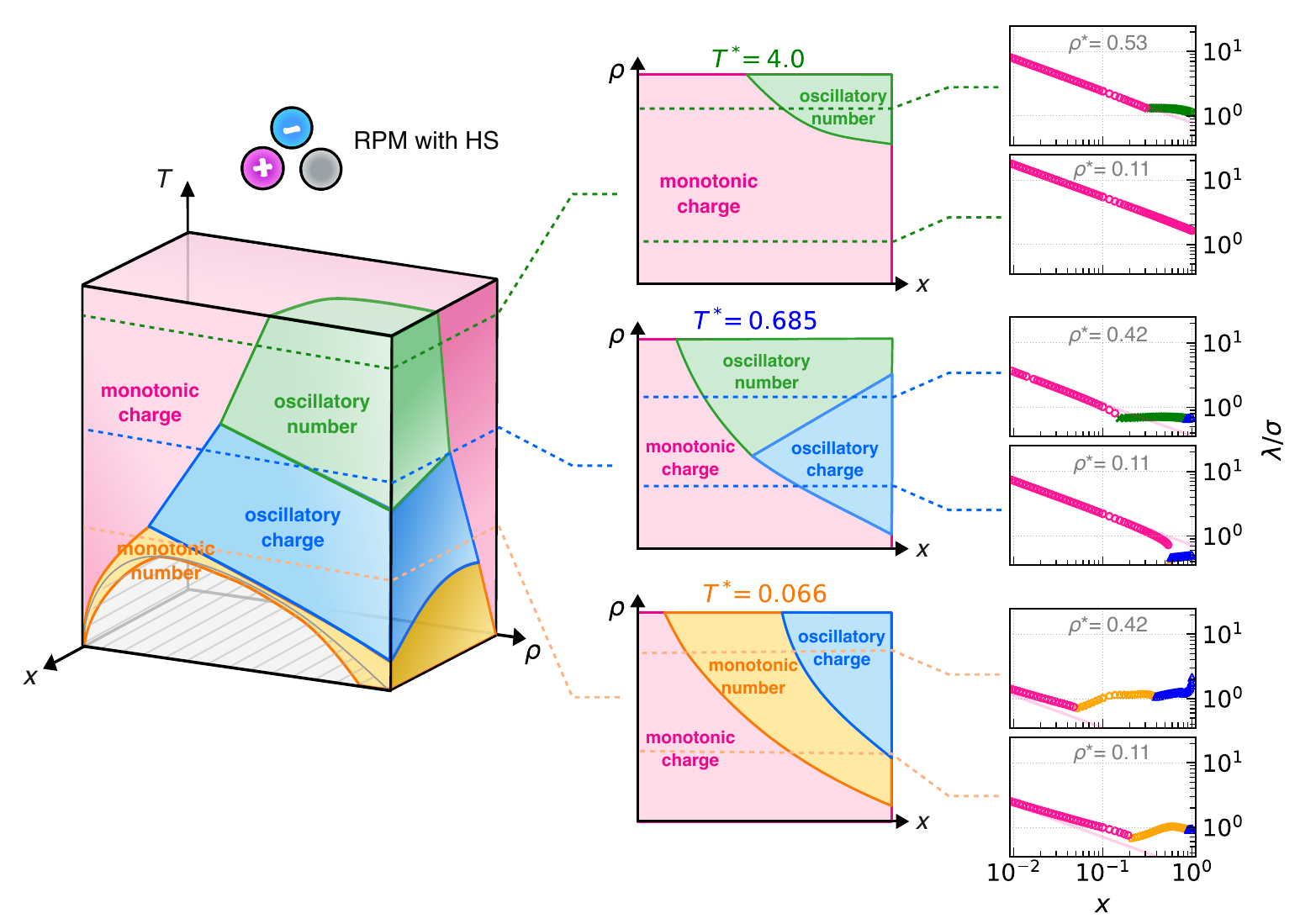}
  \caption{\tbf{Hard-sphere solvent effects.} Decay length regimes for
    the RPM with an equal sized neutral hard-sphere solvent across the
    temperature--density--concentration space (left). The middle
    column shows density--concentration cross-sections at three
    isotherms, as indicated by their labels. The right column
    shows the decay length as a function of concentration at two
    representative fixed densities per temperature, with the
      color of the symbols indicating the dominant regime.  }
\label{fig2}
\end{figure*}

The size and shape of the ions are only one source of molecular
complexity in electrolyte solutions. In solvent-containing
electrolytes, the solvent is more than a passive dielectric
background: its molecules occupy volume and can mediate directional
interactions at short range.  Experiments on ionic liquid--solvent
mixtures show that the oscillation wavelength changes with composition
\cite{Smith2017}, highlighting how competing molecular length
  scales influence packing. Explicit ion--dipole correlations can
also qualitatively alter electrolyte phase behavior
\cite{Walker2026}. Even for a neutral hard-sphere solvent, however,
existing studies \cite{Coupette2018, Yang2023} leave open how the
decay regimes evolve across the full
temperature--density--concentration space. To isolate these solvent
effects systematically, we restrict our attention to symmetric
electrolytes with $c_{++}=c_{--}$ and $c_{+0}=c_{-0}$, where the
subscript `0' denotes the solvent.

In analogy with the solvent-free case, it is convenient to work
  in a basis of ionic number density, ionic charge density, and
  solvent density. The corresponding mole-fraction-weighted
  correlation functions are
\begin{equation}
  \begin{aligned}
    c_{\rm II}(r) & = \tfrac{1}{2} x  [c_{++}(r) +c_{+-}(r) ], \\
    c_{\rm SS}(r)  & =  (1-x) c_{00}(r) , \\
    c_{\rm ZZ}(r)  &= \tfrac{1}{2} x\left[c_{++}(r)  - c_{+-}(r) \right], \\
    c_{\rm IS}(r)  &=   
     x^{1/2} (1-x)^{1/2} c_{+0}(r), 
  \end{aligned}
\end{equation}
where $x$ denotes the total ionic mole fraction. As in the RPM,
correlations between ionic charge and number density are uncoupled,
i.e., $c_{\rm IZ} = 0$. Correlations between solvent density and ionic
charge density are likewise uncoupled, i.e., $c_{\rm SZ}(r) = 0$ (see S.I.).

Following the same LMFT decomposition as the solvent-free case, we can
write the pole condition as
\begin{equation}
  \begin{aligned}
    \left[\left(1 - \rho \hat{c}_{\mathrm{R},\mathrm{II}}(\alpha) \right)\left(1 - \rho \hat{c}_{\mathrm{R},\mathrm{SS}}(\alpha) \right) - \rho^2 \hat{c}^2_{\mathrm{R},\mathrm{IS}}(\alpha)\right] \\
    \times \left[1 - \rho \hat{c}_{\mathrm{R},\mathrm{ZZ}}(\alpha)
    + \frac{\lambda_{\mathrm{D}}^{-2}}{\alpha^2}
    \exp\!\left(-\frac{\alpha^2}{4\kappa^2}\right)\right] = 0.
    \label{eqn:pole-LMFT-solvent}
  \end{aligned}
\end{equation}
Therefore, the charge poles depend only on
  $c_{\mathrm{R},\mathrm{ZZ}}$, while the number poles arise from the
coupled $2\times2$ block involving $c_{\mathrm{R},\mathrm{II}}$,
$c_{\mathrm{R},\mathrm{SS}}$ and $c_{\mathrm{R},\mathrm{IS}}$.

The simplest electrolyte with explicit solvent that we consider is the
RPM mixed with neutral hard spheres of the same diameter
$\sigma_0 = \sigma$.  Fig.~\ref{fig2} maps out the resulting decay
length behavior across the full $(T,\rho,x)$ space, with $(\rho,x)$
cross-sections at the same three $T^*$ used for the pure RPM in
Fig.~\ref{fig1}(a). It is convenient to consider behavior at fixed
total density $\rho$ since, for a given $T^\ast$, increasing $x$ then
corresponds to exchanging two solvent particles for every ion pair,
thereby keeping the total packing fraction fixed. Comparing to the
solvent-free case, we find that $x$ inherits part of $\rho$'s role in
driving the crossover between regimes by determining the ionic
strength of the solution. At $T^*=0.066$, introducing solvent has
little qualitative impact, with the same monotonic charge $\to$
monotonic number $\to$ oscillatory charge transitions observed as in
the pure RPM. At $T^\ast = 4.0$ and high densities, behavior is also
similar to the solvent-free case, with a transition from monotonic
charge to oscillatory number decay modes observed. In contrast, at
lower densities monotonic charge decay dominates for all $x$,
demonstrating that introducing explicit solvent can qualitatively
reshape long-range correlations in the fluid.

The influence of the solvent is even more pronounced at the
intermediate temperature, $T^*=0.685$. Here, we observe that the
sequence of transitions is sensitive to $\rho$. At high densities,
increasing $x$ drives the system through monotonic charge, oscillatory
number, and finally oscillatory charge decay; notably, this ordering
differs from the solvent-free RPM, where increasing density instead
drives the system through monotonic charge $\to$ oscillatory charge
$\to$ oscillatory number decay modes. Similar behaviour was reported
by Coupette \emph{et al.}  \cite{Coupette2018} At low densities, the
system instead transitions directly from monotonic charge to
oscillatory charge without an intervening oscillatory number window
\footnote{A subtlety arises in the dilute ion limit. As
  $x\rightarrow0$ at a fixed finite $\rho$, the decay length always
  recovers the Debye length, as expected for a dilute
  electrolyte. This limit is nevertheless discontinuous with the
  behavior exactly at $x=0$, as the pure hard sphere fluid has
  oscillatory (number) decay at every density
  \cite{Dijkstra2000}. Therefore, an arbitrarily small but non-zero
  ion concentration is enough to restore Debye-like monotonic
  screening, while removing the ions entirely returns the fluid to
  purely structural, oscillatory packing.}.

Irrespective of $T^\ast$, in all cases where oscillatory number poles
determine asymptotic decay, both the decay length and the period of
oscillation ($2\pi\alpha_1^{-1}$) are unchanged across $x$.
This insensitivity results from the fact that both the ions and the
solvent have the same diameter; oscillatory number decay is set purely
by the total packing fraction, independent of whether the packed
volume is occupied by charged or neutral species. Overall, the results
from this section demonstrate that the solvent's influence on the
decay length cannot be captured by density or composition
alone. Instead, the dominant asymptotic regime emerges from the
coupled effects of packing and ionic strength, which vary
independently once explicit solvent is introduced.

\subsection{Influence of the solvent molecular character: from packing
  to polarity}

\begin{figure*}[t]
  \includegraphics[width=\linewidth]{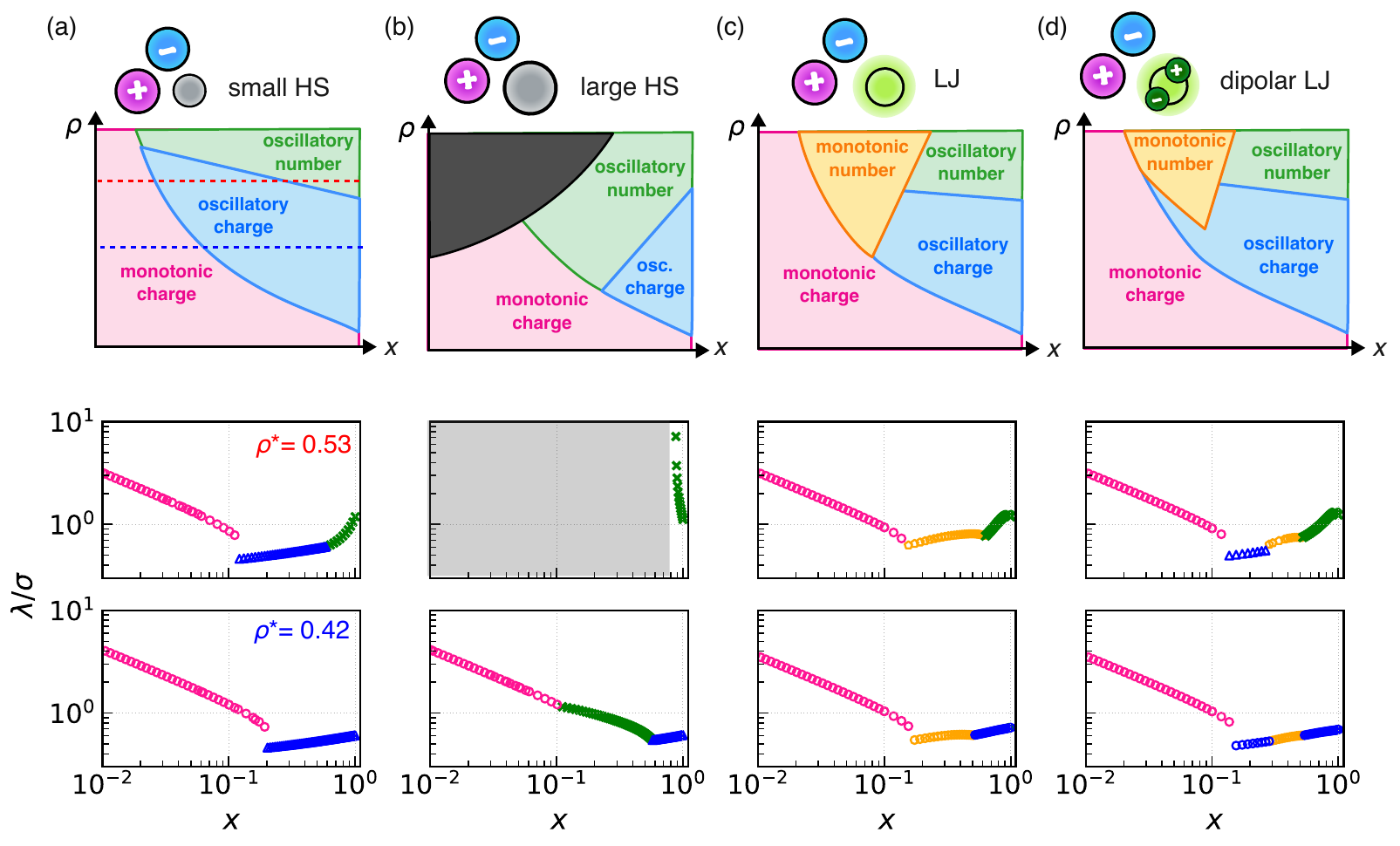}
  \caption{\tbf{Solvent molecular character reshapes the decay length
      diagram.} Decay length behavior along the isotherm $T^*=0.685$
    for the RPM with four solvent models of increasing complexity: (a)
    a small hard-sphere solvent; (b) a large hard-sphere solvent; (c)
    a Lennard-Jones solvent with soft-core repulsion and dispersion
    attraction; and (d) a dipolar LJ solvent, identical to (c) but
    carrying a permanent point dipole.  The top row shows
      schematic decay length domains in the $\rho$--$x$ plane. The
    middle and bottom rows show the decay length against mole fraction
    at fixed $\rho^*=0.53$ and $\rho^*=0.42$, respectively. The
    gray shaded regions in (b) marks as the fluid approaches its
    maximum packing fraction prior to crystallization.}
\label{fig3}
\end{figure*}

We have established that introducing a size-symmetric hard-sphere
solvent qualitatively reshapes the nature of an electrolyte's
asymptotic correlations. A natural question then arises: to what
extent does the molecular character of the solvent matter? To address
this issue, we systematically investigate how solvent properties
influence the asymptotic correlations by varying the solvent size,
softness, and intramolecular structure. For simplicity, we restrict
our investigation to cases where the ions themselves are equally sized
charged hard spheres and to a single isotherm $T^\ast\approx 0.685$.

To isolate the effect of solvent size, in Figs.~\ref{fig3}(a) and
\ref{fig3}(b) we present results from systems with a smaller
($\sigma_0=0.8\,\sigma$) and larger ($\sigma_0=1.2\,\sigma$) hard
sphere solvent, respectively. For the smaller solvent, replacing
  ion pairs with solvent displaces less excluded volume, so at any
given $(\rho,x)$, the oscillatory number decay length is weaker than
in the equal-size case.  The oscillatory charge decay mode therefore
dominates over a wider region, and only yields to oscillatory number
behavior once density becomes sufficiently high.  Consequently,
increasing $x$ at fixed density traces monotonic charge $\rightarrow$
oscillatory charge $\rightarrow$ oscillatory number, rather than the
monotonic charge $\rightarrow$ oscillatory number $\rightarrow$
oscillatory charge sequence found for the size-symmetric solvent. For
the larger solvent, the opposite holds: exchanging an ion pair for
solvent displaces proportionally more excluded volume, enhancing the
oscillatory number decay mode at fixed $(\rho,x)$. At higher
densities, the decay length of the oscillatory number mode appears to
diverge as $x$ decreases and the fluid approaches its maximum packing
fraction.

We note that, under conditions where either monotonic or oscillatory
charge modes are dominant, the magnitude of the decay length is
largely insensitive to the solvent size, which reflects the fact
  that ion diameters are the same in both systems. In stark contrast,
decay lengths associated with the oscillatory number decay are highly
sensitive to the solvent diameter, since they are controlled by the
total excluded volume of the fluid.

We aim to build toward a polar molecular solvent whose constituent
units comprise a Lennard-Jones (LJ) center and two equal-and-opposite
point charges as employed in Ref.~\cite{bui2025dielectro}. Rather than
introducing the LJ and electrostatic interactions simultaneously,
however, we isolate their effects in turn.  We take
$\sigma_{\rm LJ}\approx1.1\sigma$ and a weak interaction strength
$\varepsilon_{\rm LJ}\approx0.0054\,k_{\rm B}T$, with the LJ potential
truncated and shifted at $3.3\sigma_{\mrm{LJ}}$. Ion--solvent
interactions consist of hard-sphere repulsion with a contact distance
of $\frac{1}{2}(\sigma_{\rm LJ}+\sigma)$, supplemented by
electrostatic interactions when solvent charges are introduced below.
As can be seen in Fig.~\ref{fig3}(c) the softness of the LJ potential
means that the system more closely resembles the smaller solvent
discussed above [Fig.~\ref{fig3}(a)] rather than the larger solvent. A
qualitatively new feature, however, is the emergence of a monotonic
number decay mode at intermediate $x$, wedged between the monotonic
charge and oscillatory regimes. The absence of this regime with
hard-sphere solvents points to the solvent's attractive interactions
as the origin. This interpretation is consistent with the
Fisher--Widom line of a LJ fluid, which marks the crossover between
monotonic decay arising from attractive interactions and oscillatory
decay arising from short-range repulsions \cite{Evans1994, Evans1993,
  Fisher1969, Vega1995, Dijkstra2000}.

Finally, introducing equal-and-opposite charges on each solvent
molecule to give a dipole moment of $p^*=2$ reshapes the diagram once
more \nocite{Borgis2023, Pollock1980}
 \footnote{The reduced dipole moment is defined as
  $p^*=p/\sqrt{4\pi\epsilon_0\varepsilon_{\rm LJ}\sigma_{\rm
      LJ}^3}$. We take the parameters from earlier studies
  \cite{bui2025dielectro, Borgis2023, Pollock1980}. Under the
  conditions investigated here, the solvent has a low dielectric
  constant $\epsilon\approx1.1$}.  As seen in Fig.~\ref{fig3}(d),
comparing to the LJ only solvent, while the oscillatory-number decay
length is largely unchanged, the monotonic-number decay length
decreases, resulting in the window over which it dominates
shrinking. Consequently, we observe a re-entrant behavior at
moderately high densities: oscillatory charge $\to$ monotonic
number $\to$ oscillatory charge. These observations suggest that the
increase in ion-solvent coupling due to the solvent molecules'
permanent dipoles primarily influences the monotonic number decay
mode, leaving other modes largely unaffected.
  
The results of this section send a clear message: solvent details
matter for determining the nature of a fluid's asymptotic
correlations. Their effects are not merely quantitative: solvent
properties can also change the dominant decay mode and the sequence of
decay regimes encountered as ion concentration increases.

\subsection{The Kirkwood crossover and the limits of its universality}

The sensitivity of a fluid's asymptotic decay to the molecular
  details of both the ions and solvent argues against a universal
  description of long-ranged correlations. In the systems we have
  investigated, however, we have observed instances where the
  magnitude of monotonic and oscillatory decay lengths have remained
  largely unaffected by changes to molecular details. We therefore
  pose a more focused question: can the screening length associated
  with ionic charge correlations be described in universal terms?

\begin{figure}[t]
  \includegraphics[width=\linewidth]{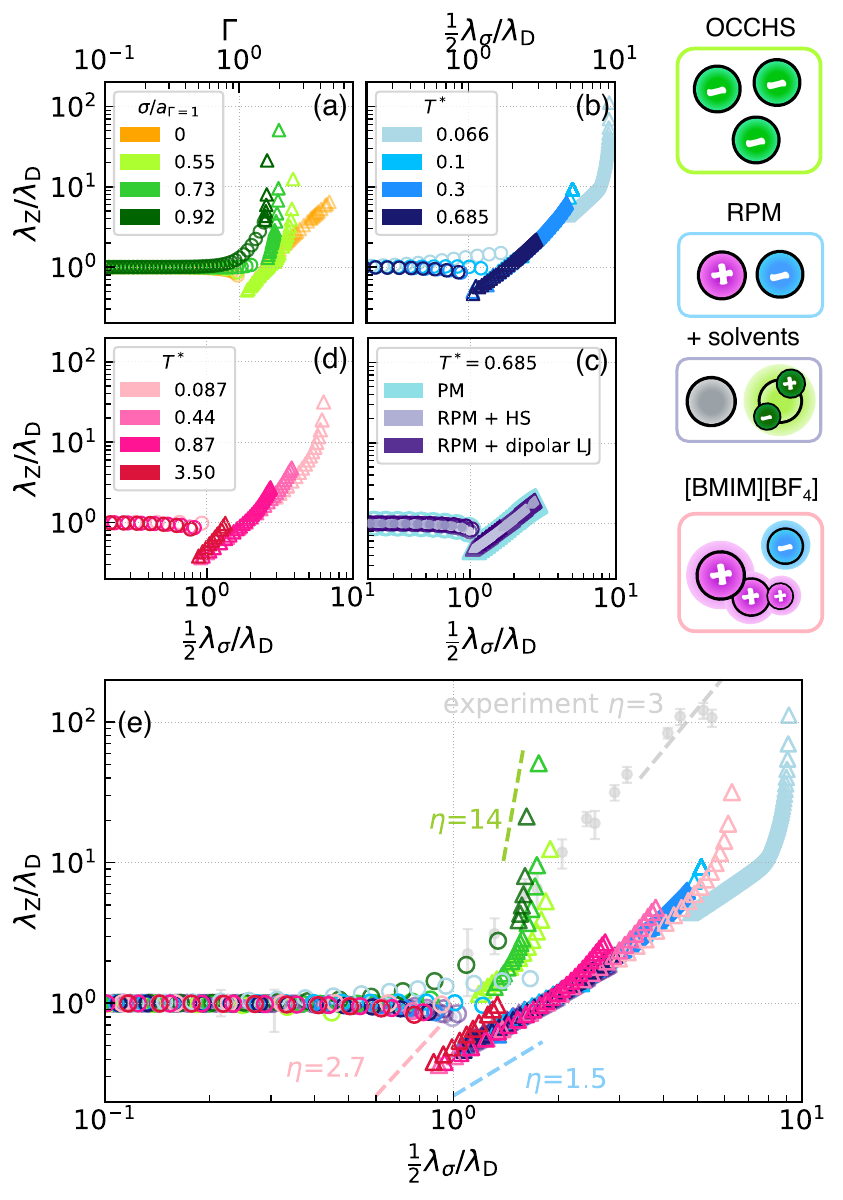}
  \caption{\tbf{The Kirkwood crossover across one-component, binary,
      and molecular ionic fluids.} Decay lengths from neural
      cDFT for (a) the OCP ($\sigma=0$) and OCCHS, (b) RPM, (c)
    coarse-grained [BMIM][BF$_4$], and (d) RPM with hard-sphere or
    dipolar solvents. The lower panel compares theoretical
      results with experimental measurements (gray) \cite{Smith2016}.
    Dashed lines indicate effective scaling exponents.} 
  \label{fig4}
\end{figure}

Across every ionic fluid considered so far, from the solvent-free RPM
to explicit molecular solvents, the monotonic charge length closely
tracks the Debye--H\"uckel prediction at low ionic strength. This is
expected on physical grounds: Debye--H\"uckel theory is a mean-field
result that becomes exact in the dilute limit regardless of
microscopic detail.  The open question does not concern this dilute
limit, but the system's fate after the Kirkwood crossover into the
oscillatory charge regime at high ion concentration. We have already
seen that this transition is not always direct: it can be bridged by
oscillatory- or monotonic-number regimes depending on the
thermodynamic state or solvent properties. We therefore focus here on
the charge modes themselves, first identifying the minimal ingredients
a system needs to possess for a Kirkwood crossover at all.

The simplest model to exhibit electrostatic screening is the one
component plasma (OCP): identical point particles interacting
through the Coulomb potential in a uniform compensating background
\cite{Hansen1973}.  Its thermodynamic properties depend on a single
dimensionless coupling parameter $\Gamma=\lambda_{\rm B}/a$, where
$a=(3/4\pi\rho)^{1/3}$ is the ion sphere radius.  In terms of its
correlations, the OCP differs from the systems considered so far in
that charge and number density fluctuations coincide.  Consequently,
the charge and number poles are identical and the pole condition
reduces to
\begin{equation}
  1 - \rho \hat{c}_{\mathrm{R}}(\alpha)
  + \frac{\lambda_{\mathrm{D}}^{-2}}{\alpha^2}
  \exp\!\left(-\frac{\alpha^2}{4\kappa^2}\right) = 0.
  \label{eqn:pole-ocp}
\end{equation}
It is well-established that the resulting decay length is
charge-monotonic at low $\Gamma$, obeying Debye--H\"uckel
theory, and transitions to charge-oscillatory at
$\Gamma\approx1$ \cite{leote_de_carvalho_decay_1999, Baus1980}, which we
reproduce with our neural cDFT approach in Fig.~\ref{fig4}(a). While
this crossover is analogous to a Kirkwood transition, the OCP lacks a
genuinely fixed packing length scale as its oscillation period is
necessarily tied to $a$, which varies continuously with density
\cite{leote_de_carvalho_decay_1999}.

To introduce genuine excluded volume packing, we consider the
one-component charged hard sphere fluid (OCCHS) \cite{Chen2004}, i.e.,
the OCP with an underlying hard-sphere core, for three hard-core
diameters $\sigma/a_{\Gamma=1}=0.55, 0.73, 0.92$.  Increasing $\sigma$
changes behavior either side of the crossover
  [Fig.~\ref{fig4}(a)]: before the crossover the monotonic decay
length deviates from the Debye length at progressively weaker
coupling, while after the crossover, the oscillatory branch grows
more steeply than in the OCP, with a period set by $\sigma$ rather
than $a$. Together, these results show that while Coulomb coupling
alone is sufficient to generate a Kirkwood crossover, excluded-volume
interactions are required to decouple the oscillation period from the
ion-sphere radius and thereby produce packing-controlled oscillatory
decay.

We are now in a position to ask whether charge correlations exhibit
universal behavior, even beyond the Kirkwood crossover. Note that this
is a weaker requirement than attempting to address the experimentally
observed anomalous underscreening; even if a set of microscopic models
appears to behave universally, such behavior may still fail to account
for the large screening lengths seen in experiments.  To this end, we
compare $\lambda_{\rm Z}/\lambda_{\rm D}$ against
$\frac{1}{2}\lambda_{\sigma}/\lambda_{\rm D}$ on a log--log scale for
all investigated models, restricting attention to regions where charge
correlations dominate. Here,
$\lambda_{\rm Z} = \max(\lambda_{\rm Z}^{\rm (mon)},\lambda_{\rm
  Z}^{\rm (osc)})$ and $\lambda_{\sigma}$ is the period of the
oscillatory charge decay mode ($\sigma$ for OCCHS, $2\sigma_{\pm}$ for
the binary systems).
  
For the RPM, the data approximately collapse beyond the Kirkwood
crossover, with an effective exponent $\eta\approx1.5$ consistent with
prior reports \cite{Jager2023, Coles2020, Rotenberg2018, Cats2021}
[Fig.~\ref{fig4}(b)]. Adding hard-sphere or dipolar solvent preserves
this behavior over the range investigated [Fig.~\ref{fig4}(c)]. This
collapse is nevertheless limited: at higher concentrations, the RPM
curves steepen, with an onset that depends on temperature. The
departure is especially pronounced at the lowest temperature studied,
near the critical point. Thus, even within the RPM, a single exponent
does not describe the entire oscillatory regime.

The other ionic models further reveal the limits of this scaling. The
coarse-grained [BMIM][BF$_4$] fluid exhibits a steeper increase than
the RPM systems especially at high temperature, followed by further
steepening at high concentration [Fig.~\ref{fig4}(d)]. For the OCCHS,
changing the hard-core diameter separates the monotonic branches,
although the oscillatory branches approximately collapse with
exponents in the range $11 \lesssim \eta \lesssim 14$
[Fig.~\ref{fig4}(a,e)]. The approximate collapse found for the RPM
therefore does not extend to a common scaling law across structurally
distinct ionic fluids. Recent work using a spread-charge mean-field model likewise finds that post-crossover screening lengths depend strongly on the ionic charge distribution \cite{Fong2026}, consistent with the sensitivity to ionic architecture observed here.

The combined comparison in Fig.~\ref{fig4}(e) places these results
alongside experimental measurements \cite{Smith2016}.  Although we do
see screening lengths of magnitude comparable to those measured
experimentally, they do so at different concentration.  Our results
suggest that the magnitude of experimental screening lengths is not in
itself anomalous: similarly large values arise in theory at high
concentrations.  The discrepancy lies instead in the apparent
universality of the experimental scaling: no single exponent describes
the ionic architectures considered here, in contrast to the common
cubic scaling reported experimentally.

\section{CONCLUSIONS}

We have used neural functional theory to probe how an ionic
  fluid's molecular details influence its long-ranged correlations,
  which are central to determining solvent-mediated interactions. By
  investigating systems of varying molecular complexity from
  solvent-free primitive models, to an ionic liquid, to systems with
  explicit solvent, we have shown that the nature of long-ranged
  correlations is sensitive to the properties of both the ions and the
  solvent. Importantly, this sensitivity is not merely quantitative:
  molecular details can alter the nature of the dominant asymptotic
  decay mode itself.

We also considered charge density correlations in isolation,
  irrespective of whether they determine the asymptotic decay of the
  fluid. While we found our results to be consistent with previous
  studies, we also found that screening lengths increased rapidly at
  sufficiently high concentrations, exceeding the Debye length by 1--2
  orders of magnitude. However, the concentrations at which we
  observed this rapid increase are dependent upon the underlying
  molecular model. This picture is difficult to reconcile with reports
  of a single universal scaling exponent from surface force
  measurements of anomalous underscreening \cite{Smith2016,
    Lee2017}. As the theoretical framework within which we have worked
  is rooted in equilibrium statistical mechanics, our results instead
  lend indirect support to more recent experimental interpretations
  that suggest nonequilibrium effects play an important role
\cite{Cross2026}.

Taken together, our results indicate that the screening length in
concentrated electrolytes should be treated as a molecular property of
the specific ion--solvent system in question, rather than assumed to
follow a single universal law.  These findings highlight the
importance of accounting for solvent packing and polarity when
interpreting screening measurements, and may help inform future
studies of concentrated electrolytes for energy storage applications.
More broadly, this work demonstrates that neural functional theory can
efficiently map screening behavior across a
temperature--density--concentration space that would be challenging to
access with conventional integral equation theories, traditional
classical density functional methods, or molecular simulations alone,
opening a path toward systematic, molecularly resolved predictions of
screening in realistic electrolytes.

\section{METHODS}

\subsection{Neural classical density functional}

We train separate neural functionals for each electrolyte
model and temperature considered in this work: the RPM at
$T^*=4.0$, $0.685$, $0.30$ and $0.066$; the asymmetric PM
at the same four temperatures; the coarse-grained
[BMIM][BF$_4$] ionic liquid at $T^*=3.50$, $0.44$ and
$0.087$; the RPM with same-sized, smaller
($\sigma_0=0.8\,\sigma$) and larger ($\sigma_0=1.2\,\sigma$)
hard-sphere solvents, a Lennard-Jones solvent, and a dipolar
LJ solvent, all at $T^*=0.685$; the OCP; and the OCCHS at
$\sigma/a_{\Gamma=1}=0.55$, $0.73$ and $0.92$.
Full simulation parameters are given in Table~S1.

For each model, a neural network is trained to represent the one-body
direct correlation functional
$c_{\mrm{R},\nu}^{(1)}(z;[\{\rho_\nu\}],T)$ of the short-ranged
reference (mimic) system defined by the LMFT splitting in
Eq.~\ref{eqn:LMFT-splitting}, following the neural functional approach
of Refs.~\cite{Sammuller2023, Sammuller2025, Bui2025prl}.  Training
data consist of 1000--2000 grand canonical Monte Carlo simulations per
state point, performed under randomised planar external potentials so
that each simulation yields a pair of density and $c^{(1)}$ profiles
across a range of inhomogeneous
environments~\cite{Sammuller2023}. Alternative learning schemes based
on pair-correlation data could also be used to obtain the reference
system two-body direct correlation functions $c_{\mrm{R},\nu\gamma}$
\cite{Dijkman2025, Kampa2026, Kampa2026b}.  For molecularly structured
species whose internal charge distribution couples number and
charge density, we instead learn a hyperdensity functional
$c_{\mrm{R},\nu}^{(1)}[\{\rho_\nu(z)\},\beta\phi(z)]$ that takes the
electrostatic potential as an additional input~\cite{bui2025hyperlmft,
  bui2025dielectro}; this applies to the BMIM cation in [BMIM][BF$_4$]
and the dipolar solvent.  Training was performed on a GPU
(\texttt{NVIDIA GeForce RTX 3060}) and completed in a few hours per
functional; evaluation of the trained functional is fast
($\sim$\,milliseconds).  Full details of the training data generation,
network architecture and hyperparameters are given in Section~S2.

The two-body direct correlation functions
$c_{\mrm{R},\nu\gamma}(z,z')$ follow by functional
differentiation of the trained $c^{(1)}_{\mrm{R},\nu}$, evaluated via
reverse-mode automatic differentiation.  Evaluating at uniform bulk
density (and zero electrostatic potential) and projecting to radial
geometry gives $c_{\mrm{R},\nu\gamma}(r)$, from which the linear
combinations entering the pole equations (Eqs.~\ref{eqn:pole-LMFT}
and~\ref{eqn:pole-LMFT-solvent}) are constructed. The radial
projection is detailed in Section~S3 of the Supporting Information.
\subsection{Pole analysis}

For each phase point, the complex poles
$\alpha=\alpha_1+i\alpha_0$ of the Ornstein--Zernike equation
are obtained by recasting the pole conditions into coupled
real-valued equations and solving numerically via a
Newton--Raphson scheme.
For symmetric ions ($c_{\mrm{NZ}}=0$), the number and charge
poles decouple. The number poles satisfy
\begin{equation}
\left\{
\begin{aligned}
\alpha_{0}  &= 4\pi \rho \!\int_0^{\infty}\!\! \mrm{d}r\, r\, c_{\mrm{R,NN}}(r)
      \sinh(\alpha_{0} r) \cos(\alpha_{1} r),\\
\alpha_{1}  &= 4\pi \rho \!\int_0^{\infty}\!\! \mrm{d}r\, r\, c_{\mrm{R,NN}}(r)
      \cosh(\alpha_{0} r) \sin(\alpha_{1} r),
\end{aligned}
\right.
\end{equation}
and the charge poles satisfy
\begin{equation}
\left\{
\begin{aligned}
 \alpha_0  - \frac{A}{\lambda^2_{\mrm{D}}}&= 4\pi \rho \! \int_0^{\infty}\!\! \mrm{d}r\, r\, c_{\mrm{R,ZZ}}(r)
      \sinh(\alpha_0 r) \cos(\alpha_1 r),\\
 \alpha_1 + \frac{B}{\lambda^2_{\mrm{D}}} &= 4\pi \rho \! \int_0^{\infty}\!\! \mrm{d}r\, r\, c_{\mrm{R,ZZ}}(r)
      \cosh(\alpha_0 r) \sin(\alpha_1 r),
\end{aligned}
\right.
\end{equation}
where $A(\alpha_0,\alpha_1)$ and $B(\alpha_0,\alpha_1)$ arise
from separating the real and imaginary parts of the Coulomb
term $\lambda_{\mrm{D}}^{-2}\alpha^{-2}\exp(-\alpha^2/4\kappa^2)$
(explicit forms are given in Section~S3).
The purely monotonic poles ($\alpha_1=0$) reduce to
one-dimensional root-finding problems in $\alpha_0$ alone.
For asymmetric ions, the coupled number--charge equations are
solved using the decoupled solutions as initial guesses, and
the same coupled procedure applies to the ion--number and
solvent--number poles in the three-component system with the
replacements $\mrm{NN}\to\mrm{II}$, $\mrm{ZZ}\to\mrm{SS}$,
$\mrm{NZ}\to\mrm{IS}$, and $A=B=0$.
All equations are reformulated in terms of the
planar-geometry direct correlation functions
$\bar{c}_{\mrm{R},\nu\gamma}(z)$, which are the direct output
of the neural functional; full expressions are given in
Section~S3.

\section*{Data Availability}

Data and code supporting the findings of this study will
be openly available upon publication of the manuscript.

\section*{Supporting Information}

Supplementary information includes simulations, machine learning details and additional supporting results.

\section*{Acknowledgements}
Via membership of the United Kingdom Car Parrinello (UKCP) consortium
(Grant No. EP/F036884/1) and of the UK’s HEC Materials Chemistry
Consortium funded by EPSRC (EP/X035859), this work used the ARCHER2 UK
National Supercomputing Service.  A.T.B. acknowledges funding from the
Oppenheimer Fund, Peterhouse College and Cambridge Philosophical Society, University of Cambridge. A.T.B. is supported by a Stanford Science Fellowship.
S.J.C. is a Royal Society University Research Fellow (Grant
No. URF\textbackslash R1\textbackslash 211144) at Durham
University.

\bibliography{references}

\end{document}


\title{Supporting information: Molecular structure breaks universality in screening behavior of electrolyte solutions}

\author{Anna T. Bui}
\email{annatbui@stanford.edu}
\affiliation{Yusuf Hamied Department of Chemistry, University of
  Cambridge, Lensfield Road, Cambridge, CB2 1EW, United Kingdom}
\affiliation{Department of Chemistry, Durham University, South Road,
  Durham, DH1 3LE, United Kingdom}
\affiliation{Department of Chemistry, Stanford University, California 94305, USA}
  
\author{Connie J. Fairchild}
\affiliation{Yusuf Hamied Department of Chemistry, University of
  Cambridge, Lensfield Road, Cambridge, CB2 1EW, United Kingdom}
  
\author{Dominic P. W. Thomas}
\affiliation{Department of Chemistry, Durham University, South Road,
  Durham, DH1 3LE, United Kingdom}
  
\author{Stephen J. Cox}
\email{stephen.j.cox@durham.ac.uk}
\affiliation{Department of Chemistry, Durham University, South Road,
  Durham, DH1 3LE, United Kingdom}

\date{\today}

\maketitle

\makeatletter
\patchcmd{\l@@sections}
  {\hskip\skip@}
  {\nobreak\kern0.5em
   \leaders\hbox{\normalfont\kern0.15em.\kern0.15em}\hfill
   \kern0.5em}
  {}
  {\PackageWarning{tocdots}{Could not add TOC dots}}
\makeatother
\tableofcontents
\newpage

\section{Theoretical framework}

We present the theoretical framework for analysing the
asymptotic decay length in ionic fluids. We start with a single
component system, before going on to multicomponent cases.

\subsection{One-component systems}

The full theory discussion is based on Ref.~\onlinecite{Evans1994}.
The bulk structure of a fluid can be characterized by the total
pair correlation function
\begin{equation}
  h(r) = g(r) - 1,
\end{equation}
where $g(r)$ is the radial distribution function, which is related to
the two-body direct correlation function $c(r)$ via the
Ornstein--Zernike equation, which, after Fourier transformation,
  reads
\begin{equation}
  \hat{h}(k) = \frac{\hat{c}(k)}{1-\rho \hat{c}(k)},
\end{equation}
where the radial Fourier transform is defined as
\begin{equation}
  \hat{f}(k) = \frac{4\pi}{k} \!\!\int^{\infty}_0\!\!\mrm{d}r\, r \sin(kr) \,f(r).
\end{equation}
Away from criticality, $c(r)\sim-\beta u(r)$, where $u(r)$ is the pair potential and
$\beta=1/(k_{\mrm{B}}T)$, with $T$ the temperature and
  $k_{\mrm{B}}$ Boltzmann's constant.  Therefore, for fluids with
  short-ranged interactions, $c(r)$ is also short-ranged and its
Fourier transform can be expanded in even powers.  Since both
$\hat{c}(k)$ and $\hat{h}(k)$ are even, we can write
\begin{equation}
  r h(r) = \frac{1}{2\pi^2} \int^\infty_0\!\!\mrm{d}k\, k \sin(k r)\hat{h}(k) =  \frac{1}{4\pi^2 i} \int^\infty_{-\infty}\!\!\mrm{d}k\, k \exp(i k r )\frac{\hat{c}(k)}{1-\rho \hat{c}(k)}.
  \label{eqn:rhform}
\end{equation}
For short-ranged fluids $\hat{h}(k)$ exhibits poles at complex
wavenumber $\alpha=\alpha_1+i\alpha_0$ satisfying
\begin{equation}
  1 - \rho \hat{c}(\alpha) = 0.
  \label{eqn:pole-condition}
\end{equation}
For fluids with short-ranged potentials, in general one
  finds
\begin{itemize}
\item A single pole on the imaginary axis
  ($\alpha = i\alpha_0^\mrm{mon}$) indicating a monotonic exponential
  decay at low density
  \begin{equation}
    r h(r) \sim \exp(-\alpha^{\mrm{mon}}_0 r),
  \end{equation}
\item A conjugate pair of poles
  $\alpha = \pm \alpha_1^{\mrm{osc}} + i \alpha^{\mrm{osc}}_0$
  indicating an exponentially damped oscillatory decay at high density
  \begin{equation}
    r h(r) \sim \cos(\alpha^{\mrm{osc}}_{1} r - \theta)\, \exp(-\alpha^{\mrm{osc}}_0 r),
  \end{equation}
  where $\alpha^{-1}_0$ denotes the decay length in each regime,
  $2\pi/\alpha^{\mrm{osc}}_{1}$ is the period of oscillation and
  $\theta$ is a phase shift.
\end{itemize}
The Fisher--Widom line marks the crossover from monotonic to
  oscillatory decay of number density correlations in the
  temperature--density plane.

For the one-component plasma and one-component charged hard
  sphere fluid, each particle carries a charge $q$. There is then a
  Coulombic contribution to the pair potential
\begin{equation}
  u(r) = \frac{q^2}{4\pi\epsilon_0\epsilon r},
\end{equation}
where $\epsilon_0$ is the permittivity of free space, and
  $\epsilon$ is the relative permittivity of the medium. We then have
  $c(r)\sim-\beta q^2/4\pi\epsilon_0\epsilon r$. 
Employing the splitting prescribed by local molecular field theory
\begin{equation}
  \frac{1}{r}= \frac{\mathrm{erfc}(\kappa r)}{r} + \frac{\mathrm{erf}(\kappa r)}{r},
\end{equation}
the direct correlation function decomposes
\begin{equation}
  c(r) = c_{\rm R}(r) - \frac{\beta q^2 \mathrm{erf}(\kappa r)}{4\pi \epsilon_0\epsilon r},
  \label{eqn:lmft-decompose}
\end{equation}
where $c_{\rm R}$ is the direct correlation function of a system
  where electrostatic interactions are mediated by the kernel
  $\mathrm{erfc}(\kappa r)/r$. As $\mathrm{erfc}(\kappa r)/r$ decays
  rapidly for $r\gg\kappa^{-1}$, $c_{\rm R}(r)$ is a short-ranged
  function.
In Fourier space, Eq.~\ref{eqn:lmft-decompose} is
\begin{equation}
  \hat{c}(k) = \hat{c}_{\mrm{R}}(k) - \frac{ \beta q^2}{\epsilon_0\epsilon k^2} \exp\left(-\frac{k^2}{4\kappa^2}\right),
\end{equation}
and the complex poles of the fluid satisfy
\begin{equation}
  1 - \rho \hat{c}_{\mrm{R}}(\alpha) + \frac{\lambda_{\mrm{D}}^{-2}}{\alpha^2} \exp\left(-\frac{\alpha^2}{4\kappa^2}\right)= 0,
\end{equation}
where we have used the Debye length
$\lambda_{\mrm{D}}=(\epsilon_0\epsilon/\beta \rho q^2)^{1/2}$.
While a short-ranged simple fluid has at most one imaginary pole, this is
no longer the case for electrostatic system due the presence of the
Coulomb term.  Generally, we find two regions:
\begin{itemize}
\item At low density, the leading pole lies on the imaginary axis at $\alpha=i\alpha^{\rm mon}_0$, giving rise to a monotonic decay,
    \begin{equation}
      r h(r) \sim \exp(-\alpha^{\mrm{mon}}_0 r).
    \end{equation}
    This pole is accompanied by another imaginary pole at
    $i\alpha_0'$, with $\alpha_0'>\alpha^{\rm mon}_0$. As the
    density increases, the poles move towards each other and coalesce.
  \item As the density is further increased, the two poles move off
    the imaginary axis and symmetrically into the complex plane
    $\alpha = \pm \alpha_1^{\mrm{osc}} + i \alpha^{\mrm{osc}}_0$,
    giving rise to oscillatory decay.
    \begin{equation}
      r h(r) \sim \cos(\alpha^{\mrm{osc}}_{1} r - \theta)\, \exp(-\alpha^{\mrm{osc}}_0 r).
    \end{equation}
\end{itemize}
This monotonic to oscillatory crossover is the Kirkwood
  transition \cite{Kirkwood1936}; it is  driven by the coalescence of
two imaginary poles, a distinct mechanism from the Fisher--Widom
crossover found in simple fluids, and is a signature of Coulombic
systems\cite{LeotedeCarvalho1994,
  leote_de_carvalho_decay_1999}. Even though the number and
  charge densities cannot be varied independently in a
  single-component fluid, it is this behavior of the poles that
  distinguishes this crossover as a Kirkwood transition.

\subsection{Two-component systems}

For a binary ionic fluid, the total and two-body direct correlation
functions, $h_{\nu\gamma}(r)$ and $c_{\nu\gamma}(r)$, are related
through the mixture Ornstein--Zernike equation which in matrix form
\cite{Baxter2003} is
\begin{equation}
 \hat{\mbf{H}}(k) = \hat{\mbf{C}}(k) + \hat{\mbf{C}}(k)\hat{\mbf{H}}(k),
\end{equation}
where the matrix elements are defined as
\begin{equation}
  \begin{split}
    \hat{H}_{\nu\gamma} (k) &= \rho(x_\nu x_\gamma)^{1/2} \hat{h}_{\nu\gamma} (k) \\ 
    \hat{C}_{\nu\gamma} (k) &= \rho(x_\nu x_\gamma)^{1/2} \hat{c}_{\nu\gamma} (k),
  \end{split}
\end{equation}
with $x_\nu$ the mole fraction of species $\nu$ and
$\rho=\sum_\nu \rho_{\nu}$ the total bulk number
density. Following analogous logic to the single component case
\cite{leote_de_carvalho_decay_1999}, the asymptotic decay is
determined by the complex poles $k=\alpha=\alpha_1 + i\alpha_0$
satisfying
\begin{equation}
  \det\!\left[\mbf{I} - \hat{\mbf{C}}(\alpha)\right] = 0.
\end{equation}
For a binary ionic fluid with equal mole fractions $x_{-}=x_{+}=1/2$,
this pole condition reduces to
\begin{equation}
\left[1 - \frac{\rho \hat{c}_{++}(\alpha)}{2}\right]
\left[1 - \frac{\rho \hat{c}_{--}(\alpha)}{2}\right]
- \frac{\rho^2 \hat{c}_{+-}^2(\alpha)}{4} = 0.
\end{equation}
Since the poles of an ionic fluid naturally separate into those
associated with number correlations and those associated with charge
correlations, it is convenient to transform from the species basis
$(+,-)$ to the number--charge $(\mathrm{N},\mathrm{Z})$ basis via the
orthogonal transformation
\begin{equation}
\hat{\mbf{C}}^{(\mathrm{N,Z})} = \mbf{w}^\intercal \hat{\mbf{C}}^{(+,-)}
\mbf{w}, \qquad
\mbf{w} =
\begin{pmatrix}
\frac{1}{\sqrt{2}} & \frac{1}{\sqrt{2}} \\
\frac{1}{\sqrt{2}} & -\frac{1}{\sqrt{2}}
\end{pmatrix}.
\end{equation}
Defining the mole-fraction-weighted correlation functions in the new basis,
where $i,j \in \{\mrm{N,Z}\}$ and $\nu,\gamma \in \{+,-\}$,
\begin{equation}
    c_{ij}(r) = \sum_{\nu,\gamma} w_{\nu i}\, w_{\gamma j}
    (x_\nu x_\gamma)^{1/2} c_{\nu\gamma}(r),
\end{equation}
so that $\hat{C}_{ij}(k)=\rho\hat{c}_{ij}(k)$, the transformation yields
the following linear combinations for a binary ionic fluid
\begin{equation}
\begin{aligned}
    c_{\mrm{NN}}(r) & = \frac{1}{4} \left[c_{++}(r) + c_{--}(r) + 2 c_{+-}(r) \right], \\
    c_{\mrm{ZZ}}(r) &=  \frac{1}{4} \left[ c_{++}(r) +  c_{--}(r)  - 2 c_{+-}(r) \right], \\
    c_{\mrm{NZ}}(r) &=  \frac{1}{4} \left[  c_{++}(r) -   c_{--}(r) \right], 
\end{aligned}
\end{equation}
and analogous expressions for $h_{\nu\gamma}(r)$.
In this basis the pole condition takes the form
\begin{equation}
\left[1 - \rho \hat{c}_{\mrm{NN}}(\alpha)\right]
\left[1 - \rho \hat{c}_{\mrm{ZZ}}(\alpha)\right]
- \rho^2 \hat{c}_{\mrm{NZ}}^2(\alpha) = 0.
\label{eqn:pole-NZ-basis}
\end{equation}
With the LMFT splitting,
\begin{equation}
  c_{\nu\gamma}(r) = c_{\rm R, \nu\gamma}(r) - \frac{\beta q_{\nu}  q_{\gamma} \mathrm{erf}(\kappa r)}{4\pi \epsilon_0\epsilon r}
    \label{eqn:lmft-decompose-2component}
\end{equation}
we can write
\begin{equation}
  \begin{split}
    c_{\rm NN}(r) &= \frac{1}{4}  \left[c_{\mrm{R}, ++}(r)   + c_{\mrm{R}, --}(r) + 2c_{\mrm{R},+-}(r)\right]  \equiv c_{\mrm{R,NN}}(r)\\
    c_{\rm ZZ}(r) &=  \frac{1}{4}  \left[c_{\mrm{R}, ++}(r) + c_{\mrm{R}, --}(r)  - 2c_{\mrm{R},+-}(r) - \frac{\beta  q^2\mathrm{erf}(\kappa r)}{\pi \epsilon_0\epsilon r}\right]   \equiv c_{\mrm{R,ZZ}}(r)- \frac{\beta q^2\mathrm{erf}(\kappa r)}{4\pi\epsilon_0\epsilon r}  \\
    c_{\rm NZ}(r) &= \frac{1}{4}\left[ c_{ \mrm{R}, ++}(r) - c_{\mrm{R}, --}(r) \right] \equiv  c_{\mrm{R, NZ}}(r),
  \end{split}
\end{equation}
where we have used $q_{+}=-q_{-}=q$.
Equation~\ref{eqn:pole-NZ-basis} then reads
\begin{equation}
\begin{aligned}
\left[1 - \rho \hat{c}_{\mrm{R, NN}}(\alpha)\right]
\left[1 - \rho \hat{c}_{\mathrm{R},\mathrm{ZZ}}(\alpha)
+ \frac{\lambda_{\mathrm{D}}^{-2}}{\alpha^2}
\exp\!\left(-\frac{\alpha^2}{4\kappa^2}\right)\right] 
 - \rho^2 \hat{c}_{\mrm{R,NZ}}^2(\alpha)= 0.
\label{eqn:pole-LMFT}
\end{aligned}
\end{equation}
For symmetric ions where $\hat{c}_{\rm R, NZ}(\alpha)=0$, one can  find two sets of number-dominated poles from
\begin{equation}
    1 - \rho\hat{c}_{\rm R, NN}(\alpha) = 0, 
\end{equation}
and two set of charge-dominated poles from 
\begin{equation}
1 - \rho \hat{c}_{\mathrm{R},\mathrm{ZZ}}(\alpha)
+ \frac{\lambda_{\mathrm{D}}^{-2}}{\alpha^2}
\exp\!\left(-\frac{\alpha^2}{4\kappa^2}\right) =0.
\end{equation}
Here, the Debye length is still
$\lambda_{\mrm{D}}=(\epsilon_0\epsilon/\beta \rho q^2)^{1/2}$,
and we recall that $\rho$ is the total number density.  One can
find up to four types of poles
\begin{itemize}
    \item Monotonic number: with a pole lying on the imaginary axis at $\alpha =  i \alpha^{\mrm{mon}}_{0,\mrm{N}}$ giving rise to an exponentially monotonic decay, 
    \begin{equation}
        r h_{++}(r) \sim r h_{+-}(r) \sim 
     \exp(-\alpha^{\mrm{mon}}_{0,\mrm{N}} r ).
    \end{equation}
    \item Oscillatory number: with conjugate pairs of poles  $\alpha = \pm \alpha_{1,\mrm{N}}^{\mrm{osc}}  + i \alpha^{\mrm{osc}}_{0,\mrm{N}}$ giving rise to oscillatory decay,
    \begin{equation}
      r h_{++}(r) \sim r h_{+-}(r) \sim \cos(\alpha^{\mrm{osc}}_{1,\mrm{N}} r - \theta_{\mrm{N}})\, \exp(-\alpha^{\mrm{osc}}_{0,\mrm{N}} r).
    \end{equation}
     \item Monotonic charge: with with a pole lying on the imaginary axis at $\alpha =  i \alpha^{\mrm{mon}}_{0,\mrm{Z}}$ giving rise to an exponentially monotonic decay, 
    \begin{equation}
      r h_{++}(r) \sim -r h_{+-}(r) 
        \sim 
     \exp(-\alpha^{\mrm{mon}}_{0,\mrm{Z}} r ).
    \end{equation}
    This pole is accompanied by another imaginary pole at
    $i\alpha_{0,\mrm{Z}}'$, with
    $\alpha_{0,\mrm{Z}}'>\alpha^{\rm mon}_{0,\mrm{Z}}$. As the
    density increases, the poles move towards each other and coalesce.
    \item Oscillatory charge: with conjugate pairs of poles  $\alpha = \pm \alpha_{1,\mrm{Z}}^{\mrm{osc}}  + i \alpha^{\mrm{osc}}_{0,\mrm{Z}}$ giving rise to oscillatory decay,
    \begin{equation}
      r h_{++}(r) \sim -r h_{+-}(r)  \sim \cos(\alpha^{\mrm{osc}}_{1,\mrm{Z}} r - \theta_{\mrm{Z}})\, \exp(-\alpha^{\mrm{osc}}_{0,\mrm{Z}} r).
    \end{equation}
\end{itemize}

\subsection{Three-component systems}

For a binary electrolyte with explicit solvent, we transform
from the species basis $(+,-,0)$ to the $(\mrm{I,S,Z})$ basis where
the I denotes the ion number, S denotes the solvent number, and Z
denotes the charge, via
\begin{equation}
\hat{\mbf{C}}^{(\mrm{I,S,Z})} = \mbf{w}^\intercal \hat{\mbf{C}}^{(+,-, \mrm{s})} \mbf{w}, \qquad
\mbf{w} =
\begin{pmatrix} 
 \frac{1}{\sqrt{2}} &  0 &  \frac{1}{\sqrt{2}} \\
 \frac{1}{\sqrt{2}} & 0 & -\frac{1}{\sqrt{2}} \\
0 & 1 & 0 
\end{pmatrix}.
\end{equation}
In general, these give the linear combinations
\begin{equation}
\begin{aligned}
    c_{\rm II}(r) & = \frac{1}{2}\left[x_{+} c_{++}(r) + x_{-} c_{--}(r) 
    + 2 (x_{+} x_{-})^{1/2} c_{+-}(r)  \right], \\
    c_{\rm SS}(r) & = x_0  c_{00}(r), \\
c_{\rm ZZ}(r) &=  \frac{1}{2}\left[x_{+} c_{++}(r) +  x_{-} c_{--}(r)  - 2(x_{+} x_{-})^{1/2} c_{+-}(r)\right], \\
c_{\rm IZ}(r) &= \frac{1}{2} \left[x_+  c_{++}(r) - x_-  c_{--}(r)\right], \\
    c_{\rm IS}(r) &= \frac{1}{\sqrt{2}}\left[ (x_+x_0)^{1/2} c_{+0}(r) + (x_-x_0)^{1/2} c_{-0}(r)  \right], \\
c_{\rm SZ}(r) &= \frac{1}{\sqrt{2}} \left[ (x_+x_0)^{1/2}  c_{+0}(r)
- (x_-x_0)^{1/2}  c_{-0}(r) \right].
\end{aligned}
\end{equation}
For symmetric ions with equal and opposite charges
($c_{++} = c_{--}$, and $c_{+0} = c_{-0}$, and
$x_+ = x_- \equiv \frac{1}{2}x$, where $x$ is the total ion mole
fraction), these simplify to
\begin{equation}
\begin{aligned}
    c_{\rm II}(r) & = \frac{1}{2} x  [c_{++}(r) +c_{+-}(r) ], \\
    c_{\rm SS}(r)  & =  (1-x) c_{00}(r) , \\
    c_{\rm ZZ}(r)  &= \frac{1}{2} x\left[c_{++}(r)  - c_{+-}(r) \right], \\
    c_{\rm IS}(r)  &=   
     x^{1/2} (1-x)^{1/2} c_{+0}(r) , \\
    c_{\rm IZ}(r)  &= 0, \\
    c_{\rm SZ}(r)  &= 0.
\end{aligned}
\end{equation}
For symmetric ions, the charge correlations decouple from the ion-number and solvent correlations, so that the pole condition factorises into
\begin{equation}
    \left[\left(1 - \rho\hat{c}_{\mrm{II}}(\alpha)\right)\left(1 - \rho\hat{c}_{\mrm{SS}}(\alpha)\right) - \rho^2 \hat{c}_{\mrm{IS}}^2(\alpha)\right] \left[1 - \rho \hat{c}_{\mrm{ZZ}}(\alpha) \right]  = 0.
    \label{eqn:pole-ISZ-basis}
\end{equation}
Applying the LMFT decomposition as prescribed in Eq.~\ref{eqn:lmft-decompose-2component},
\begin{equation}
\begin{aligned}
    c_{\rm II}(r) & = \frac{1}{2} x  [c_{\mrm{R}, ++}(r) + c_{\mrm{R}, +-}(r) ] \equiv c_{\rm R, II}(r), \\
    c_{\rm SS}(r)  & =  (1-x) c_{\mrm{R}, 00}(r) \equiv c_{\rm R, SS}(r), \\
    c_{\rm ZZ}(r)  &= \frac{1}{2} x\left[c_{\mrm{R}, ++}(r)  - c_{\mrm{R}, +-}(r) - \frac{\beta  q^2\mathrm{erf}(\kappa r)}{2\pi \epsilon_0\epsilon r}\right] \equiv c_{\rm R, ZZ}(r) - \frac{\beta x q^2 \mathrm{erf}(\kappa r)}{4\pi\epsilon_0\epsilon r}, \\
    c_{\rm IS}(r)  &=   
     x^{1/2} (1-x)^{1/2} c_{\mrm{R},+0}(r) \equiv c_{\rm R, IS}(r), \\
    c_{\rm IZ}(r)  &= 0, \\
    c_{\rm SZ}(r)  &= 0,
\end{aligned}
\end{equation}
allows us to rewrite Eq.~\ref{eqn:pole-ISZ-basis} as
\begin{equation}
\left[\left(1 - \rho \hat{c}_{\mathrm{R},\mathrm{II}}(\alpha) \right)\left(1 - \rho \hat{c}_{\mathrm{R},\mathrm{SS}}(\alpha) \right) - \rho^2 \hat{c}^2_{\mathrm{R},\mathrm{IS}}(\alpha)\right] 
 \left[1 - \rho \hat{c}_{\mathrm{R},\mathrm{ZZ}}(\alpha)
+ \frac{\lambda_{\mathrm{D}}^{-2}}{\alpha^2}
\exp\!\left(-\frac{\alpha^2}{4\kappa^2}\right)\right] = 0.
\label{eqn:pole-LMFT-solvent}
\end{equation}
The charge-dominated poles are determined by the same condition as in
the solvent-free binary case, with the Debye length
$\lambda_{\mrm{D}}=(\epsilon_0\epsilon/\beta x \rho q^2)^{1/2}$ now also
accounting for the ion mole fraction. The number-dominated poles,
however, are determined by the coupled ion-number and solvent
correlations through the first factor of
Eq.~\ref{eqn:pole-LMFT-solvent}, and thus depend on the ion--solvent
coupling encoded in $c_{\mathrm{R},\mathrm{IS}}$.

\section{Neural cDFT training}

\subsection{Generation of training data}

For the RPM (both with and without hard sphere solvent), PM, OCP
  and OCCHS, the one-body direct correlation function of each species
in the mimic system can be written exactly as
\begin{equation}
    c^{(1)}_{\rm R, \nu}(z) = \ln\left( \Lambda_{\nu}^3 \rho_{\rm R, \nu}(z)\right) + \beta V_{\rm R, \nu}(z) + \beta q_\nu \phi_{\rm R}(z) - \beta \mu_{\rm R, \nu},
    \label{eqn:c1-formula}
\end{equation}
where $\Lambda_{\nu}$, $V_{\rm R, \nu}$, $q_\nu$ and
$\mu_{\rm R, \nu}$ are the de Broglie thermal wavelength, external
non-electrostatic potential, charge and chemical potential of species
$\nu$, respectively, and $\phi_{\rm R}$ is the external
electrostatic potential.  To learn the functional relationship
$c^{(1)}_{\rm R, \nu}(z;[\{\rho_{\rm R, \nu}\}], T)$ for each
electrolyte, we perform 1000--2000 grand canonical Monte Carlo (GCMC)
simulations with randomized values of $\{\mu_{\rm R, \nu}\}$ combined
with planar external electrostatic and non-electrostatic potentials of
randomized magnitude.  Following Ref.~\onlinecite{Sammuller2023}, the
external non-electrostatic potentials take the form
\begin{equation}
    V_{\mrm{R},\nu}(z) = \sum^4_{n=1} a_n \sin\left( \frac{2\pi n z}{L_{z}} + \theta_n \right) + \sum_n b^{\mrm{lin}}_n(z),
\end{equation}
where $a_n$ and $\theta_n$ are randomly selected Fourier coefficients
and phases, respectively, and $L_z$ is the simulation box length in
the $z$ direction.  The phases $\theta_n$ are chosen uniformly in the
interval $[0, 2\pi)$, and the amplitudes $a_n$ are drawn from a normal
distribution.  The second summation
denotes up to four piecewise linear functions
\begin{equation}
    b^{\mrm{lin}}_n(z) = \begin{cases}
    V_1 + \dfrac{V_2 - V_1}{z_2 - z_1}(z - z_1) & \quad z_1 < z < z_2\\
     0 & \quad \text{otherwise},
   \end{cases}
\end{equation}
with $0 < z_1 < z_2 < L_z$.  The locations $z_1$ and $z_2$ are
randomly chosen uniformly in the interval $(0,L_z)$, while $V_1$
and $V_2$ are drawn from a normal distribution with zero mean and
variance $(k_{\mrm{B}}T)^2$.  The external electrostatic potential is
randomized analogously 
\begin{equation}
    \phi_{\rm R}(z) = \sum^4_{n=1} b^{q}_n \sin\left( \frac{2\pi n z}{L_{z}} + \theta^q_n \right) + \sum_n b^{\mrm{lin},q}_n(z).
\end{equation}
Cubic simulation boxes with periodic boundary conditions were used,
and the thermal wavelengths were set to $\Lambda_\nu=1\,\mrm{\AA}$ for
all species.  GCMC simulations were performed with our own
code~\cite{githubgcmc}.  Since we probe specific isotherms, we train
separate functionals at each fixed $T$ rather than treating
  temperature as an input to the neural network \cite{Sammuller2025,
  bui2025dielectro}.  For each simulation, the density profile of each
species was sampled on a grid with spacing $\Delta z=0.03\,\mrm{\AA}$,
and the corresponding one-body direct correlation profiles were
calculated from Eq.~\ref{eqn:c1-formula}.  Table~\ref{tab:systems}
summarises the parameters for all systems studied.

\begin{table}[H]
    \centering
    \begin{tabular}{l l}
    \hline
    \hline
    System & Parameters in real units  \\
    \hline
     RPM ($T^*=4.0,\, 0.685,\, 0.30,\, 0.066$) 
     & $\sigma_{+}=\sigma_{-}=\sigma=2.76\,\mrm{\AA}$, $q_{+}=-q_{-}=e$, \\
     & $\kappa^{-1}=5\,\mrm{\AA}$, $T=242424,\, 41515,\, 18182,\, 4000$\,K   \\[4pt] 
     PM ($T^*=4.0,\, 0.685,\, 0.30,\, 0.066$) 
     & $\sigma_{+}=1.84\,\mrm{\AA}$, $\sigma_{-}=3.68\,\mrm{\AA}$, $q_{+}=-q_{-}=e$, \\ 
     & $\kappa^{-1}=5\,\mrm{\AA}$, $T=242424,\, 41515,\, 18182,\, 4000$\,K \\[4pt] 
     {[}BMIM{]}{[}BF$_4${]} ($T^*=3.50,\, 0.44,\, 0.087$) 
     & As given in Ref.~\onlinecite{Merlet2012}, \\
     & $\kappa^{-1}=8\,\mrm{\AA}$, $T=80000,\, 10000,\, 2000$\,K \\[4pt]
     RPM + same-sized HS ($T^*=0.685$) 
     & ions as RPM, $T=41515$\,K, $\kappa^{-1}=5\,\mrm{\AA}$, \\
     & $\sigma_{0}=2.76\,\mrm{\AA}$ \\[4pt] 
     RPM + smaller HS ($T^*=0.685$) 
     & ions as RPM, $T=41515$\,K, 
     $\kappa^{-1}=5\,\mrm{\AA}$, \\ & $\sigma_{0}=2.208\,\mrm{\AA}$\\[4pt] 
     RPM + larger HS ($T^*=0.685$) 
     & ions as RPM, $T=41515$\,K, 
     $\kappa^{-1}=5\,\mrm{\AA}$, \\ & $\sigma_{0}=3.312\,\mrm{\AA}$\\[4pt] 
     RPM + LJ ($T^*=0.685$) 
     & ions as RPM, $T=41515$\,K, $\sigma_{\mrm{LJ}}=3.024\,\mrm{\AA}$, \\
     & $\varepsilon_{\mrm{LJ}}=1.87\,\mrm{kJ\,mol^{-1}}$ \\[4pt] 
     RPM + dipolar LJ ($T^*=0.685$) 
     & ions as RPM, $T=41515$\,K, LJ solvent with dipole \\
     & from two charges of $\pm 0.382\,e$ separated by $1\,\mrm{\AA}$ \\[4pt] 
     OCP 
     & $q/e=-0.14$, $T=300$\,K, $\kappa^{-1}=10\,\mrm{\AA}$\\[4pt] 
     OCCHS ($\sigma/a_{\Gamma=1}=0.55,\, 0.73,\, 0.92$) 
     & $q/e=-0.14$, $T=300$\,K, $\kappa^{-1}=10\,\mrm{\AA}$, \\
     & $\sigma=6,\, 8,\, 10\,\mrm{\AA}$  \\
    \hline
    \hline
    \end{tabular}
    \caption{\textbf{Neural functionals trained.} All RPM, PM and HS
      systems use the cross-interaction diameter
      $\sigma_{\nu\gamma}=(\sigma_{\nu}+\sigma_{\gamma})/2$. We have used $\epsilon = 1$ throughout.  }
\label{tab:systems}
\end{table}

\subsection{Learning neural density functionals}

For each fluid at each temperature, we train a neural network to
represent $c_{\mrm{R}}^{(1)}[\{\rho_{\nu}(z)\}]$ following the local
learning strategy of Ref.~\onlinecite{Sammuller2023}.  The input
consists of the density profile of each species within a sliding
spatial window extending $10\,\mrm{\AA}$ on either side of the
position of interest.  The network contains three fully
connected layers with 128, 64 and 32 nodes, respectively, each with
softplus activation.  Models were trained for 100 epochs with a batch
size of 128 using the Adam optimiser~\cite{kingma2017} and an
exponentially decaying learning rate starting at 0.001, achieving
errors comparable to the estimated simulation noise.  One fifth of the
dataset was reserved for validation.  The machine learning routine was
implemented in Keras/TensorFlow~\cite{Chollet2017}, and training was
performed on a GPU (\texttt{NVIDIA GeForce RTX 3060}) in a few hours.
Evaluating the trained neural functional is fast
($\sim$\,milliseconds) and can be performed on a single CPU core or
GPU.

\subsection{Hyperdensity functional extensions}

For fluids with molecular structure, such as the coarse-grained model
for the ionic liquid [BMIM][BF$_4$] and the dipolar LJ solvent, the
coupling between mass and charge density within a single molecular
species gives rise to electrostriction that cannot be captured by a
functional of the density alone.  To account for this, we learn a
``hyperdensity functional''
$c_{\mrm{R}}^{(1)}([\{\rho_{\rm R, \nu}(z)\},
  \beta\phi_{\mrm{R}}(z)])$ that takes both the density profiles and
the electrostatic potential as inputs~\cite{bui2025hyperlmft,
  bui2025dielectro}.  For species carrying an internal charge
distribution, the one-body direct correlation function is obtained
from
\begin{equation}
    c^{(1)}_{\rm R, \nu}(z) = \ln\left( \Lambda_{\nu}^3 \rho_{\rm R, \nu}(z)\right) + \beta V_{\rm R, \nu}(z) - \beta \mu_{\rm R, \nu},
    \label{eqn:c1-formula-hyper}
\end{equation}
where the electrostatic contribution is absent because it is absorbed into the functional dependence on $\phi_{\rm R}(z)$.
This applies to the BMIM cation in [BMIM][BF$_4$] and the dipolar solvent in the RPM + dipolar LJ system.
The neural network architecture and training procedure are otherwise identical to those described above, with the input layer extended to include $\beta\phi_{\rm R}(z)$ within the same $10\,\mrm{\AA}$ spatial window.

\section{Numerically solving for poles with neural cDFT}

We now describe the numerical procedure for solving the complex pole conditions. The direct correlation functions entering these conditions are obtained from a neural classical density functional, and it is convenient to work in real space.

\subsection{Direct correlation functions from neural density functional theory}

The short-ranged direct correlation functions $c_{\nu\gamma}(r)$ and their linear combination that enter the pole equations can be obtained from a machine-learned classical density functional.  In the neural functional approach \cite{Sammuller2023, Sammuller2025, Bui2025prl}, the one-body direct
correlation functional
\begin{equation}
    c_{\mrm{R},\nu}^{(1)}(z;[\{\rho_{\nu}\}], T) = -\frac{\delta\beta\mcl{F}_{\mrm{ex, R}}([\{\rho_{\nu}\}], T)}{\delta \rho_{\nu}(z)},
\end{equation}
where $\mcl{F}_{\mrm{ex, R}}$ is the excess Helmholtz free energy functional of the reference fluid,
is trained on grand canonical Monte Carlo data in planar geometry.  The
two-body direct correlation function then follows by functional differentiation,
\begin{equation}
     c_{\mrm{R},\nu\gamma}(z,z'; [\{\rho_{\nu}\}], T) = \frac{\delta  c_{\mrm{R},\nu}^{(1)}(z;[\{\rho_{\nu}\}],T) }{\delta \rho_{\gamma}(z')}
\end{equation}
evaluated via reverse-mode automatic differentiation.  The bulk direct
correlation function in planar geometry is the special case $\bar{c}_{\mrm{R},\nu\gamma}(z)=c_{\mrm{R},\nu\gamma}(z,z'=0;[\{\rho_{\nu, \mrm{b}}\}], T)$, which relates to its counterpart in radial geometry through
\begin{equation}
    c_{\mrm{R},\nu\gamma}(r)=- \left.\frac{1}{2\pi r} \frac{\mrm{d} \bar{c}_{\mrm{R},\nu\gamma}(z) }{\mrm{d} z}\right|_{z=r}.
    \label{eqn:c2-planar-relation}
\end{equation}

\subsection{Number-dominated poles}

For number-dominated poles, which satisfy a pole condition of the form
\begin{equation}
1 - \rho \hat{c}_{\mathrm{R},\mathrm{NN}}(\alpha) = 0, 
\end{equation}
by 
separating real and imaginary parts for
$\alpha = \alpha_{1} + i\alpha_{0}$ we can write the coupled equations
\begin{equation}
\left\{
\begin{aligned}
\alpha_{0}  &= 4\pi \rho \!\int_0^{\infty}\!\! \mrm{d}r\, r c_{\mrm{R,NN}}(r)
      \sinh(\alpha_{0} r) \cos(\alpha_{1} r),\\
\alpha_{1}  &= 4\pi \rho \!\int_0^{\infty}\!\! \mrm{d}r\, r c_{\mrm{R,NN}}(r)
      \cosh(\alpha_{0} r) \sin(\alpha_{1} r).
\end{aligned}
\right.
\label{eqn:simultaneous-pole-number}
\end{equation}
To write in terms of  the planar geometry quantity
$\bar{c}_{\mrm{R},\nu\gamma}(z)$ as the direct output of the neural
functional, one can use Eq.~\ref{eqn:c2-planar-relation} and
substitute into Eq.~\eqref{eqn:simultaneous-pole-number}. Integration by parts then allows the
integrals to be rewritten as 
\begin{equation}
\left\{
\begin{aligned}
\alpha_0   = 2 \rho \!\int_0^{\infty}\!\! \mrm{d}z\, \bar{c}_{\mrm{R},\mrm{NN}}(z)&\big[\alpha_0 \cosh(\alpha_0 z ) \cos(\alpha_1 z)\big.  \big.- \alpha_1 \sinh(\alpha_0 z ) \sin(\alpha_1 z)\big]
     ,\\
\alpha_1   = 2 \rho \!\int_0^{\infty}\!\! \mrm{d}z\, \bar{c}_{\mrm{R},\mrm{NN}}(z)
     & \big[ \alpha_0\sinh(\alpha_0 z ) \sin(\alpha_1 z) \big.  \big.+  \alpha_1\cosh(\alpha_0 z ) \cos(\alpha_1 z)\big],
\end{aligned}
\right.
\end{equation}
Solving these coupled equations numerically via a Newton--Raphson
scheme yields the number oscillatory poles
$(\alpha_{0,\mrm{N}}^{\mrm{osc}},\alpha_{1,\mrm{N}}^{\mrm{osc}})$.
The purely imaginary monotonic poles are determined by the simpler
conditions
\begin{equation}
    1   = 2 \rho \! \int^\infty_0 \!\!\mrm{d}z\, \bar{c}_{\mrm{R,NN}}(z) \cosh(\alpha_{0,\mrm{N}}^{\mrm{mon}} z ).
\end{equation}
The procedure thus yields up to two characteristic decay lengths that are number-dominated,
\begin{equation}
\begin{split}
    &\lambda^{\mrm{mon}}_{\mrm{N}} = (\alpha_{\mrm{0, N}}^{\mrm{mon}})^{-1}, \quad\quad \lambda^{\mrm{osc}}_{\mrm{N}}  = (\alpha_{\mrm{0, N}}^{\mrm{osc}})^{-1}. 
\end{split}
\end{equation}

\subsection{Charge-dominated poles}

Similarly, for  charge-dominated poles  which satisfy
 \begin{equation}
     1 - \rho \hat{c}_{\mathrm{R},\mathrm{ZZ}}(\alpha)
+ \frac{\lambda_{\mathrm{D}}^{-2}}{\alpha^2}
\exp\!\left(-\frac{\alpha^2}{4\kappa^2}\right) = 0,
\label{eqn:charge-poles}
 \end{equation}
by 
separating real and imaginary parts for
$\alpha = \alpha_{1} + i\alpha_{0}$ we can write the coupled equations
\begin{equation}
\left\{
\begin{aligned}
 & \alpha_0  - \frac{A(\alpha_0,\alpha_1)}{\lambda^2_{\mrm{D}}}= 4\pi \rho \! \int_0^{\infty}\!\! \mrm{d}r\, r c_{\mrm{R},\mrm{ZZ}}(r)
      \sinh(\alpha_0 r) \cos(\alpha_1 r),\\
 & \alpha_1 + \frac{B(\alpha_0,\alpha_1)}{\lambda^2_{\mrm{D}}} = 4\pi \rho \! \int_0^{\infty}\!\! \mrm{d}r\, r c_{\mrm{R},\mrm{ZZ}}(r)
      \cosh(\alpha_0 r) \sin(\alpha_1 r),
\end{aligned}
\right.
\label{eqn:simultaneous-pole-charge}
\end{equation}
where
\begin{equation}
    \begin{split}
        A(\alpha_0,\alpha_1) &= \left[\frac{\alpha_0}{|\alpha|^2} \cos\!\left(\frac{\alpha_1\alpha_0}{2\kappa^2}\right) + \frac{\alpha_1}{|\alpha|^2} \sin\!\left(\frac{\alpha_1\alpha_0}{2\kappa^2}\right)\right] \!\exp\!\left(\frac{\alpha_0^2-\alpha_1^2}{4\kappa^2}\right)  \\
         B(\alpha_0,\alpha_1) &= \left[\frac{\alpha_1}{|\alpha|^2} \cos\!\left(\frac{\alpha_1\alpha_0}{2\kappa^2}\right) -\frac{\alpha_0}{|\alpha|^2} \sin\!\left(\frac{\alpha_1\alpha_0}{2\kappa^2}\right)\right]\!\exp\!\left(\frac{\alpha_0^2-\alpha_1^2}{4\kappa^2}\right) 
    \end{split}
    \label{eqn:define-AB}
\end{equation}
with $|\alpha|^2=\alpha^2_1 + \alpha^2_0$. Writing in terms of  the planar geometry quantity,
\begin{equation}
\left\{
\begin{aligned}
\alpha_0  - \frac{A(\alpha_0,\alpha_1)}{\lambda^2_{\mrm{D}}} = 2 \rho \!\int_0^{\infty}\!\! \mrm{d}z\, \bar{c}_{\mrm{R},\mrm{ZZ}}(z)&\big[\alpha_0 \cosh(\alpha_0 z ) \cos(\alpha_1 z)\big.  \big.- \alpha_1 \sinh(\alpha_0 z ) \sin(\alpha_1 z)\big]
     ,\\
\alpha_1 + \frac{B(\alpha_0,\alpha_1)}{\lambda^2_{\mrm{D}}}  = 2 \rho \!\int_0^{\infty}\!\! \mrm{d}z\, \bar{c}_{\mrm{R},\mrm{ZZ}}(z)
     & \big[ \alpha_0\sinh(\alpha_0 z ) \sin(\alpha_1 z) \big. \big.+  \alpha_1\cosh(\alpha_0 z ) \cos(\alpha_1 z)\big].
\end{aligned}
\right.
\end{equation}
Solving these coupled equations yields the charge oscillatory poles
$(\alpha_{0,\mrm{Z}}^{\mrm{osc}},\alpha_{1,\mrm{Z}}^{\mrm{osc}})$.  The purely imaginary monotonic poles $\alpha=i\alpha_{0,\mrm{Z}}^{\mrm{mon}}$ are
determined by the simpler condition
\begin{equation}
    1 - \frac{\lambda_{\mrm{D}}^{-2}}{ (\alpha_{0,\mrm{Z}}^{\mrm{mon}})^2} \exp\!\left(\frac{(\alpha_{0,\mrm{Z}}^{\mrm{mon}})^2}{4\kappa^2}\right)  = 2 \rho  \!\int^\infty_0 \!\!\mrm{d}z\, \bar{c}_{\mrm{R},\mrm{ZZ}}(z) \cosh(\alpha_{0,\mrm{Z}}^{\mrm{mon}} z).
\end{equation}
The procedure thus yields up to two characteristic decay lengths that are charge-dominated,
\begin{equation}
\begin{split}
 \lambda^{\mrm{mon}}_{\mrm{Z}}  = (\alpha_{\mrm{0, Z}}^{\mrm{mon}})^{-1}, \quad\quad \lambda^{\mrm{osc}}_{\mrm{Z}}  = (\alpha_{\mrm{0, Z}}^{\mrm{osc}})^{-1}.
\end{split}
\end{equation}

\subsection{Coupled poles}

For asymmetric ions, the number and charge correlations are coupled. The poles are determined by
\begin{equation}
\left[ 1 - \rho\,\hat{c}_{\rm R, NN}(\alpha) \right]\,
\left[ 1 - \rho\,\hat{c}_{\rm R, ZZ}(\alpha) + \frac{\lambda_{\rm D}^{-2}}{\alpha^2}\exp\!\left(-\frac{\alpha^2}{4\kappa^2}\right) \right]
-\rho^2\left(\hat{c}_{\rm R, NZ}(\alpha)\right)^2 = 0.
\end{equation}
We define the shorthand
\begin{equation}
    \begin{split}
    S_{ij} &
=4\pi\rho\!\int_0^\infty\!\mrm{d}r\, r c_{\mrm{R}, ij}(r) \sinh(\alpha_0 r)\cos(\alpha_1 r),\\
    C_{ij} &
=4\pi\rho\!\int_0^\infty\!\mrm{d}r\, r c_{\mrm{R}, ij}(r) \cosh(\alpha_0 r)\sin(\alpha_1 r). 
    \end{split}
\end{equation}
In terms of the planar correlation functions, these become
\begin{equation}
    \begin{split}
    S_{ij} &
=2\rho \!\int_0^{\infty}\!\! \mrm{d}z\,  \bar{c}_{\mrm{R}, ij}(z=r) \left[ \alpha_0\cosh(\alpha_0 z ) \cos(\alpha_1 z) - \alpha_1\sinh(\alpha_0 z ) \sin(\alpha_1 z)\right],\\
    C_{ij} &
= 2\rho \!\int_0^{\infty}\!\! \mrm{d}z\,  \bar{c}_{\mrm{R}, ij}(z=r)\left[\alpha_0 \sinh(\alpha_0 z ) \sin(\alpha_1 z) + \alpha_1 \cosh(\alpha_0 z ) \cos(\alpha_1 z)\right].
    \end{split}
\end{equation}
Since \(\rho\hat{c}_{\mrm{R},ij}=(C_{ij}+ iS_{ij})/\alpha\), multiplying the pole condition by \(\alpha^2\)  and separating real and imaginary parts gives
\begin{equation}
\left\{
\begin{aligned}
(\alpha_1 &-  C_{\rm NN}) \left(\alpha_1 - C_{\rm ZZ} + \frac{B}{\lambda^2_{\mrm{D}}}\right)  - 
(\alpha_0 -  S_{\rm NN})\left(\alpha_0 -  S_{\rm ZZ} - \frac{A}{\lambda^2_{\mrm{D}} }\right) - \left(C^2_{\rm NZ} - S^2_{\rm NZ}\right)   = 0  \\[4pt]
(\alpha_1 &- C_{\rm NN}) \left(\alpha_0 -  S_{\rm ZZ} - \frac{A}{\lambda^2_{\mrm{D}} }\right)  + (\alpha_0 -  S_{\rm NN}) \left(\alpha_1 -  C_{\rm ZZ} + \frac{B}{\lambda^2_{\mrm{D}}}\right) - 2C_{\rm NZ}  S_{\rm NZ} = 0.
\end{aligned}
\right.
\end{equation}
where $A(\alpha_0, \alpha_1)$ and $B(\alpha_0, \alpha_1)$ are previously defined in Eq.~\ref{eqn:define-AB}.
Pure imaginary poles (\(\alpha_1=0\)) satisfy 
\begin{equation}
(\alpha_0 - S_{\rm NN})\left(\alpha_0 - S_{\rm ZZ} - \frac{ \exp\left(\alpha_0^2/4\kappa^2\right)}{\alpha_0 \lambda_{\mathrm D}^2}\right)-S_{\rm NZ}^2=0.
\end{equation}
We solve these numerically via a Newton--Raphson scheme, using the decoupled poles
($\hat{c}_{\mrm{R, NZ}}=0$) as initial guesses.

The same coupled procedure applies to the number-dominated poles in the three-component system, where the first factor of Eq.~\ref{eqn:pole-LMFT-solvent} couples the ion--number and solvent--number correlations. 
The equations are identical in form with the replacements of NN with II, ZZ with SS, NZ with IS, and with the Coulomb terms absent ($A=B=0$).

\section{Additional results}

\begin{figure}[H]
\centering
  \includegraphics[width=0.5\linewidth]{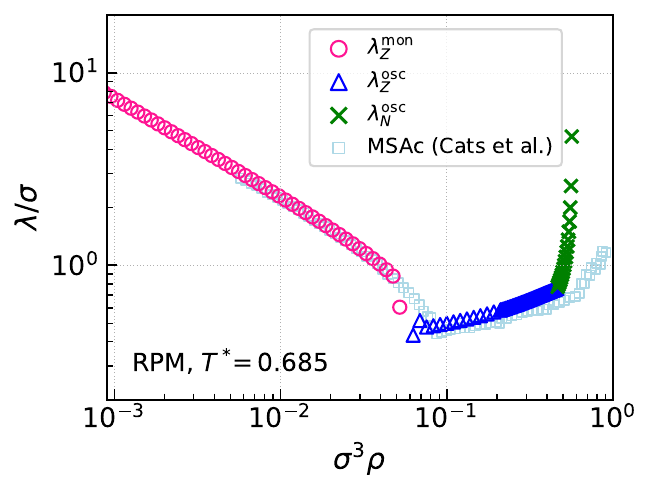}
  \caption{\textbf{Comparison of decay lengths of the RPM at
      $T^\ast = 0.685$ between neural cDFT and the MSAc functional.}
    At each density, only the longest decay length is shown, with the
    symbol indicating its character: monotonic charge-density decay
    (pink circles), oscillatory charge-density decay (blue triangles),
    and oscillatory number-density decay (green crosses). Light-blue
    squares show the corresponding results from MSAc functional
    reported in Ref.~\onlinecite{Cats2021}.}
  \label{figS1}
\end{figure}

Figure~\ref{figS1} compares the dominant decay lengths obtained from
neural cDFT with those reported for the MSAc functional in
Ref.~\onlinecite{Cats2021} at $T^\ast = 0.685$. Neural cDFT and MSAc
show close agreement for the monotonic and oscillatory charge decay
lengths. While the oscillatory number decay lengths are larger with
neural cDFT than MSAc, the crossover density is in fair
agreement. Importantly, the sequence of decay regimes predicted by
both approaches, and the transitions between them, are in overall good
agreement.

\bibliography{references}